\documentclass[twoside,twocolumn,9pt]{article}
\usepackage{amsfonts, amsmath, amssymb, mathrsfs, mathtools, amsthm, bbm, bm}
\usepackage{braket}
\usepackage{extsizes}
\usepackage[super,sort&compress,comma]{natbib} 
\usepackage[version=3]{mhchem}
\usepackage[left=1.5cm, right=1.5cm, top=1.785cm, bottom=2.0cm]{geometry}
\usepackage{balance}
\usepackage{sectsty}
\usepackage{graphicx} 
\usepackage{lastpage}
\usepackage[format=plain,justification=justified,singlelinecheck=false,font={stretch=1.125,normalsize,sf},labelfont=bf,labelsep=space]{caption}
\usepackage[font=footnotesize,justification=justified]{subcaption}
\usepackage{float}
\usepackage{fancyhdr}
\usepackage{fnpos}
\usepackage[english]{babel}
\addto{\captionsenglish}{%
  \renewcommand{\refname}{Notes and references}
}
\usepackage{tikz}
\usetikzlibrary{shadows,arrows,positioning,backgrounds,fit,shapes}
\usepackage{array}
\usepackage{droidsans}
\usepackage{charter}
\usepackage[T1]{fontenc}
\usepackage{xcolor}
\usepackage{setspace}
\usepackage[compact]{titlesec}

\usepackage{epstopdf}

\usepackage{dirtytalk} 
\usepackage{amsmath}
\usepackage{physics} 

\usepackage{color} 

\usepackage{hyperref}

\definecolor{cream}{RGB}{222,217,201}

\begin{document}

\pagestyle{fancy}
\thispagestyle{plain}
\fancypagestyle{plain}{
\renewcommand{\headrulewidth}{0pt}
}

\makeFNbottom
\makeatletter
\renewcommand\LARGE{\@setfontsize\LARGE{15pt}{17}}
\renewcommand\Large{\@setfontsize\Large{12pt}{14}}
\renewcommand\large{\@setfontsize\large{10pt}{12}}
\renewcommand\footnotesize{\@setfontsize\footnotesize{7pt}{10}}
\makeatother

\renewcommand{\thefootnote}{\fnsymbol{footnote}}
\renewcommand\footnoterule{\vspace*{1pt}%
\color{cream}\hrule width 3.5in height 0.4pt \color{black}\vspace*{5pt}} 
\setcounter{secnumdepth}{5}

\makeatletter 
\renewcommand\@biblabel[1]{#1}            
\renewcommand\@makefntext[1]%
{\noindent\makebox[0pt][r]{\@thefnmark\,}#1}
\makeatother 
\renewcommand{\figurename}{\small{Fig.}~}
\sectionfont{\sffamily\Large}
\subsectionfont{\normalsize}
\subsubsectionfont{\bf}
\setstretch{1.125} 
\setlength{\skip\footins}{0.8cm}
\setlength{\footnotesep}{0.25cm}
\setlength{\jot}{10pt}
\titlespacing*{\section}{0pt}{4pt}{4pt}
\titlespacing*{\subsection}{0pt}{15pt}{1pt}

\fancyfoot{}
\fancyfoot[LO,RE]{\vspace{-7.1pt}\includegraphics[height=9pt]{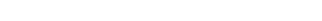}}
\fancyfoot[CO]{\vspace{-7.1pt}\hspace{13.2cm}\includegraphics{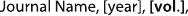}}
\fancyfoot[CE]{\vspace{-7.2pt}\hspace{-14.2cm}\includegraphics{head_foot/RF}}
\fancyfoot[RO]{\footnotesize{\sffamily{1--\pageref{LastPage} ~\textbar  \hspace{2pt}\thepage}}}
\fancyfoot[LE]{\footnotesize{\sffamily{\thepage~\textbar\hspace{3.45cm} 1--\pageref{LastPage}}}}
\fancyhead{}
\renewcommand{\headrulewidth}{0pt} 
\renewcommand{\footrulewidth}{0pt}
\setlength{\arrayrulewidth}{1pt}
\setlength{\columnsep}{6.5mm}
\setlength\bibsep{1pt}

\makeatletter 
\newlength{\figrulesep} 
\setlength{\figrulesep}{0.5\textfloatsep} 

\newcommand{\topfigrule}{\vspace*{-1pt}%
\noindent{\color{cream}\rule[-\figrulesep]{\columnwidth}{1.5pt}} }

\newcommand{\botfigrule}{\vspace*{-2pt}%
\noindent{\color{cream}\rule[\figrulesep]{\columnwidth}{1.5pt}} }

\newcommand{\dblfigrule}{\vspace*{-1pt}%
\noindent{\color{cream}\rule[-\figrulesep]{\textwidth}{1.5pt}} }

\makeatother

\twocolumn[
  \begin{@twocolumnfalse}
{\includegraphics[height=30pt]{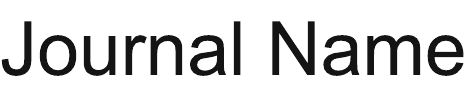}\hfill\raisebox{0pt}[0pt][0pt]{\includegraphics[height=55pt]{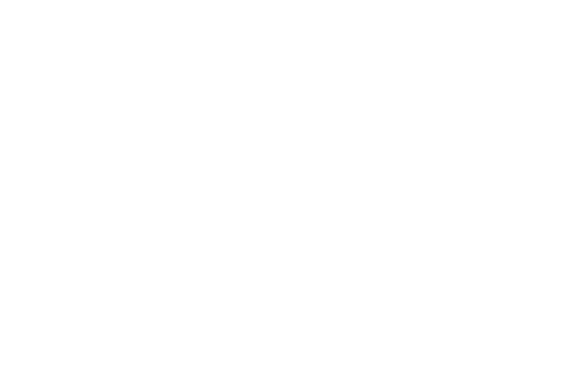}}\\[1ex]
\includegraphics[width=18.5cm]{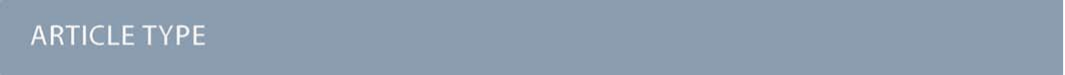}}\par
\vspace{1em}
\sffamily
\begin{tabular}{m{4.3cm} p{13.7cm} }

\includegraphics{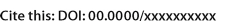} & \noindent\LARGE{\textbf{Extending Quantum Computing through Subspace, \mbox{Embedding} and Classical Molecular Dynamics Techniques}} \\
\vspace{0.3cm} & \vspace{0.3cm} \\

 & \noindent\large{Thomas M. Bickley,\textit{$^{\dag a}$} 
    Angus Mingare,\textit{$^{\dag a}$}
    Tim Weaving,\textit{$^{\dag a}$}}
    Michael Williams de la Bastida,\textit{$^{\dag a}$}
    Shunzhou Wan,\textit{$^{a}$} 
    Martina Nibbi,\textit{$^{b}$} 
    Philipp Seitz,\textit{$^{b}$} 
    Alexis Ralli,\textit{$^{ac}$} 
    Peter J. Love,\textit{$^{cd}$} 
    Minh Chung,\textit{$^{e}$} 
    Mario Hern\'andez Vera,\textit{$^{e}$} 
    Laura Schulz\textit{$^{\ddag e}$} 
    and Peter V. Coveney\textit{$^{*afg}$} \\

\includegraphics{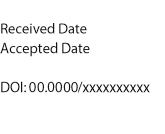} & \noindent\normalsize{The advent of hybrid computing platforms consisting of quantum processing units integrated with conventional high-performance computing brings new opportunities for algorithm design. By strategically offloading select portions of the workload to classical hardware where tractable, we may broaden the applicability of quantum computation in the near term. In this perspective, we review techniques that facilitate the study of subdomains of chemical systems with quantum computers and present a proof-of-concept demonstration of quantum-selected configuration interaction deployed within a multiscale/multiphysics simulation workflow leveraging classical molecular dynamics, projection-based embedding and qubit subspace tools. This allows the technology to be utilised for simulating systems of real scientific and industrial interest, which not only brings true quantum utility closer to realisation but is also relevant as we look forward to the fault-tolerant regime.}

\end{tabular}

 \end{@twocolumnfalse} \vspace{0.6cm}

  ]

\renewcommand*\rmdefault{bch}\normalfont\upshape
\rmfamily
\section*{}
\vspace{-1cm}


\footnotetext{\textit{$^{\ast}$~E-mail: p.v.coveney@ucl.ac.uk}}
\footnotetext{\textit{$^{\dag}$} These authors contributed equally to this work.}
\footnotetext{\textit{$^{a}$~Centre for Computational Science, Department of Chemistry, University College London, London WC1H 0AJ, United Kingdom}}
\footnotetext{\textit{$^{b}$~Technical University of Munich, School of Computation, Information and Technology, Department of Computer Science, Boltzmannstraße 3, 85748 Garching, Germany}}
\footnotetext{\textit{$^{c}$~Department of Physics and Astronomy, Tufts University 574 Boston Avenue, Medford, MA 02155, USA}}
\footnotetext{\textit{$^{d}$~Computational Science Initiative, Brookhaven National Laboratory, Upton, NY 11973, USA}}
\footnotetext{\textit{$^{e}$~Leibniz Supercomputing Centre of the Bavarian Academy of Sciences and Humanities, Boltzmannstraße 1, 85748 Garching, Germany}}
\footnotetext{\textit{$^{f}$~UCL Centre for Advanced Research Computing, Gower Street, London WC1E 6BT, United Kingdom}}
\footnotetext{\textit{$^{g}$~Informatics Institute, University of Amsterdam, Amsterdam 1098 XH, Netherlands}}
\footnotetext{\textit{$^{\ddag}$~Current address: Argonne National Laboratory, 9700 S Cass Ave, Lemont, IL 60439, USA}}



\section{Introduction}

\tikzset{back group/.style={fill=cyan!10,rounded corners, draw=black!50, dashed, inner xsep=12pt, inner ysep=12pt},}

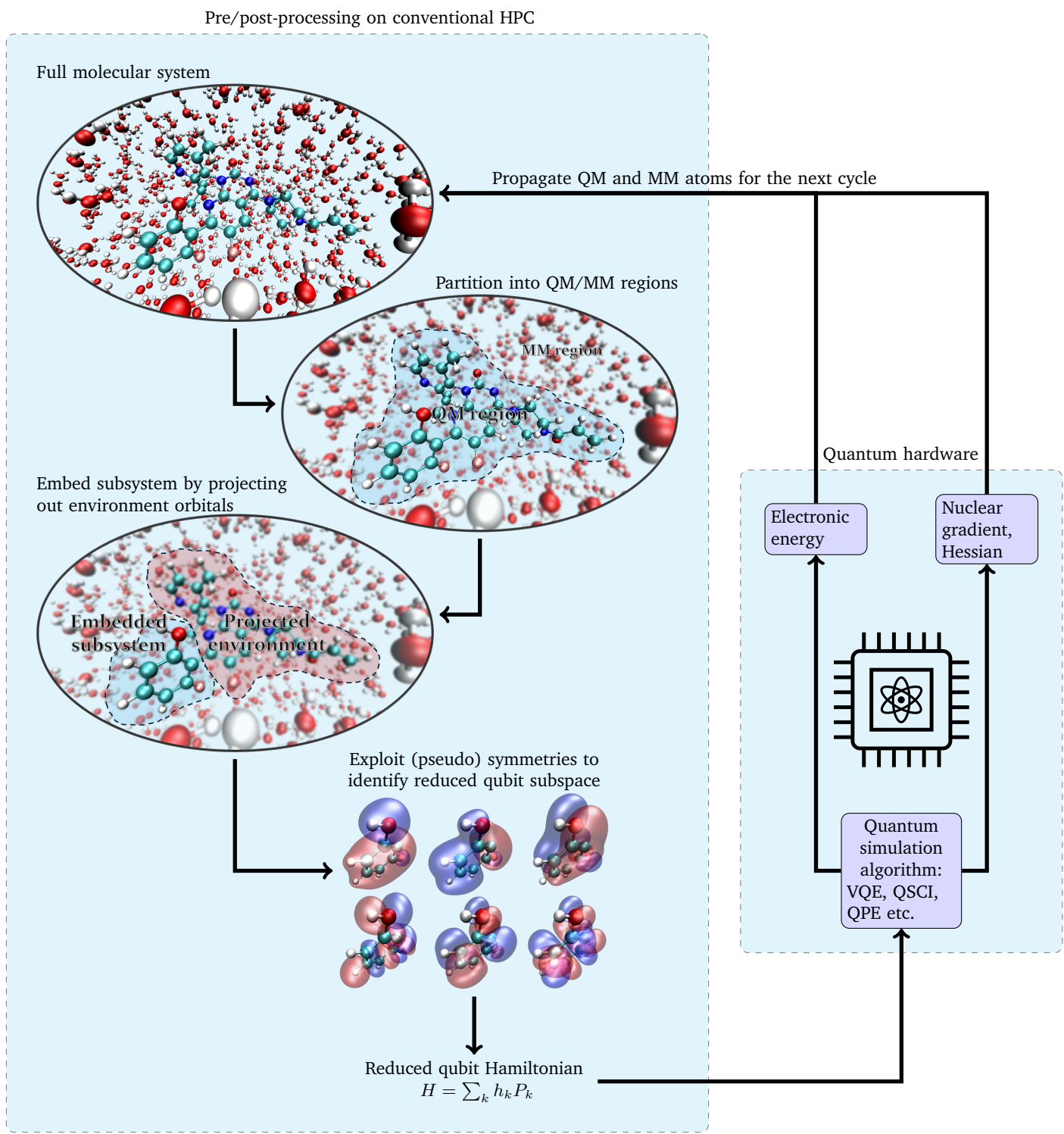
\begin{figure*}
    \centering
    \begin{tikzpicture}[node distance=3.7cm, text width=22em]
        \node (full)
            {\raggedright Full molecular system \\ \includegraphics[width=\linewidth]{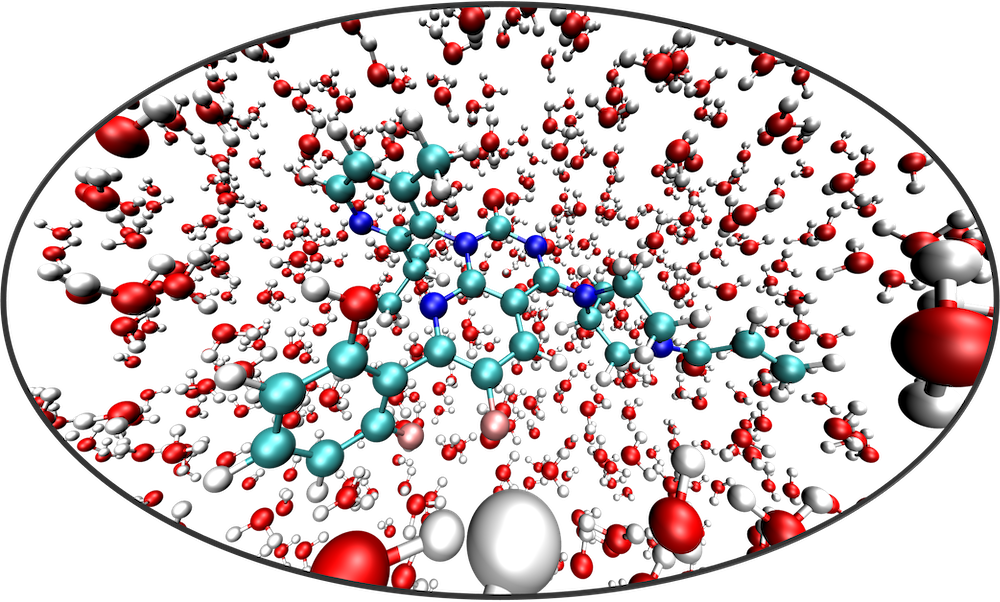}};
        \node (_qmmm) [below of=full] {}; \node (qmmm) [right of=_qmmm, node distance=4.3cm] 
            {\raggedleft Partition into QM/MM regions \\ \includegraphics[width=\linewidth]{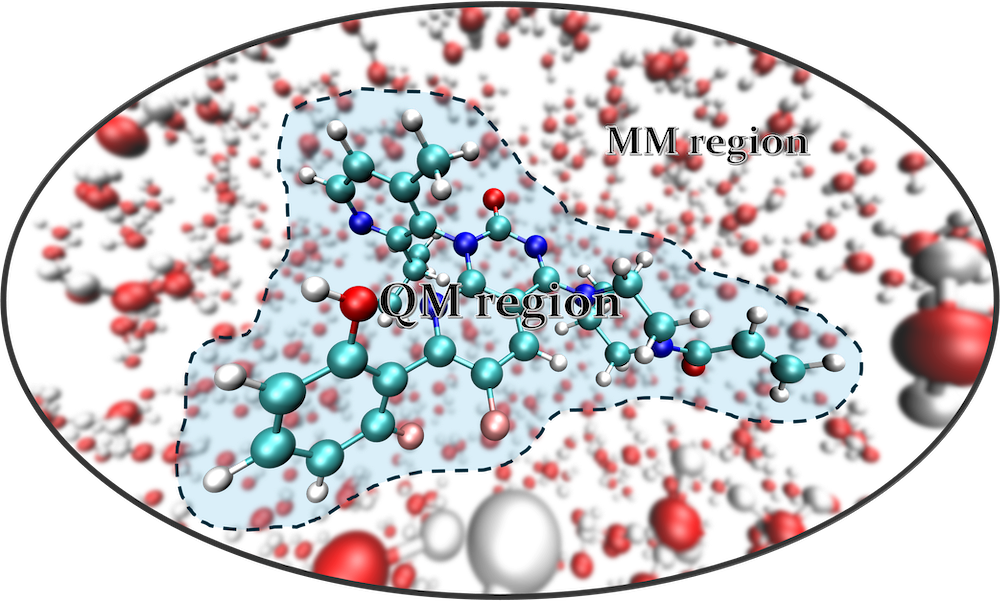}};
        \node (_embed) [below of=qmmm] {}; \node (embed) [left of=_embed, node distance=4.3cm] 
            {\raggedright Embed subsystem by projecting\\ out environment orbitals \\ \includegraphics[width=\linewidth]{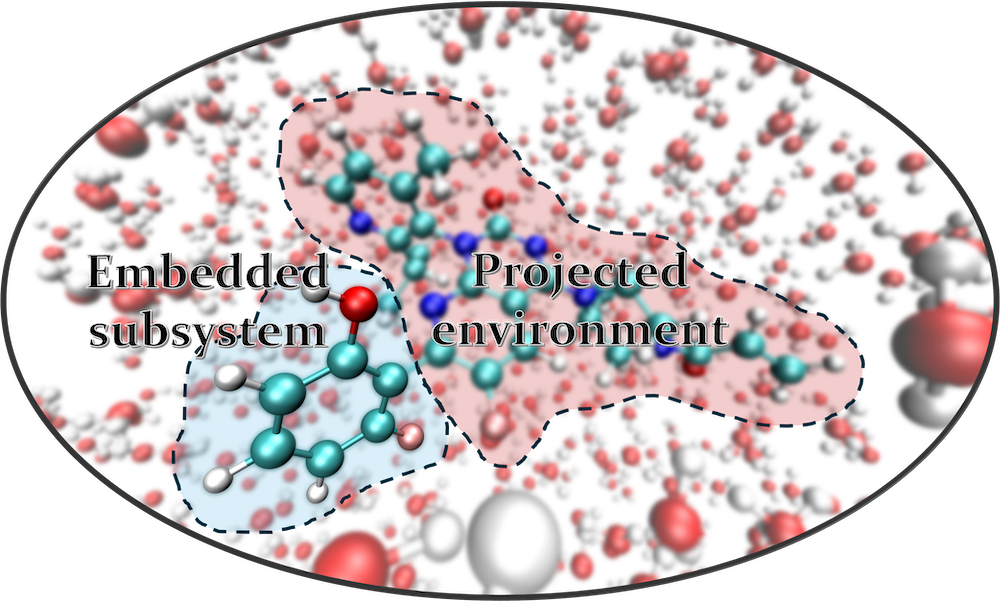}};
        \node (_orbs) [below of=embed, node distance=4.5cm] {}; \node (orbs) [right of=_orbs, text width=15em, node distance=4.2cm]
            {\centering Exploit (pseudo) symmetries to identify reduced qubit subspace  \\ \includegraphics[width=\linewidth]{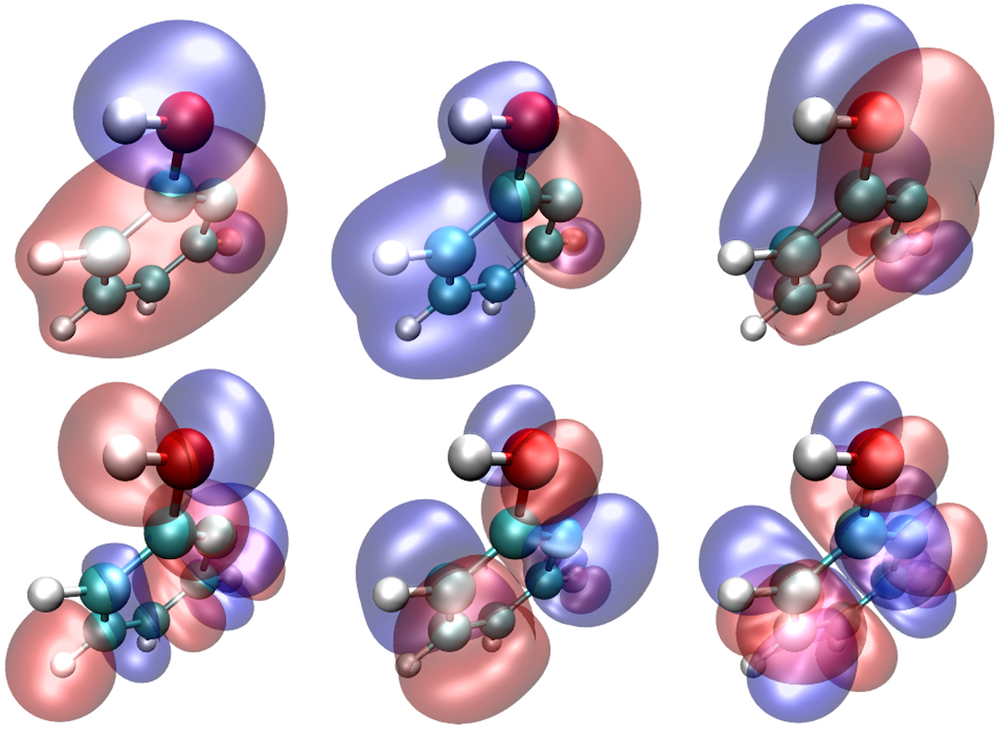}};
        \node[text width=13em] (ham) [below of=orbs]
            {\centering Reduced qubit Hamiltonian \\ \hspace{1.0cm} $H = \sum_{k}h_kP_k$};
        \node (_qsim) [right of=ham, node distance=7.5cm] {}; \node (qsim) [draw, rounded corners, fill=blue!15, above of=_qsim, text width=6em]
            {\centering Quantum simulation algorithm: \\ VQE, QSCI, QPE etc.};
        \node (_anchor) [above of=qsim, node distance=6cm] {};
        \node [below of=_anchor, node distance=3cm, text width=10em] {\includegraphics[width=\linewidth]{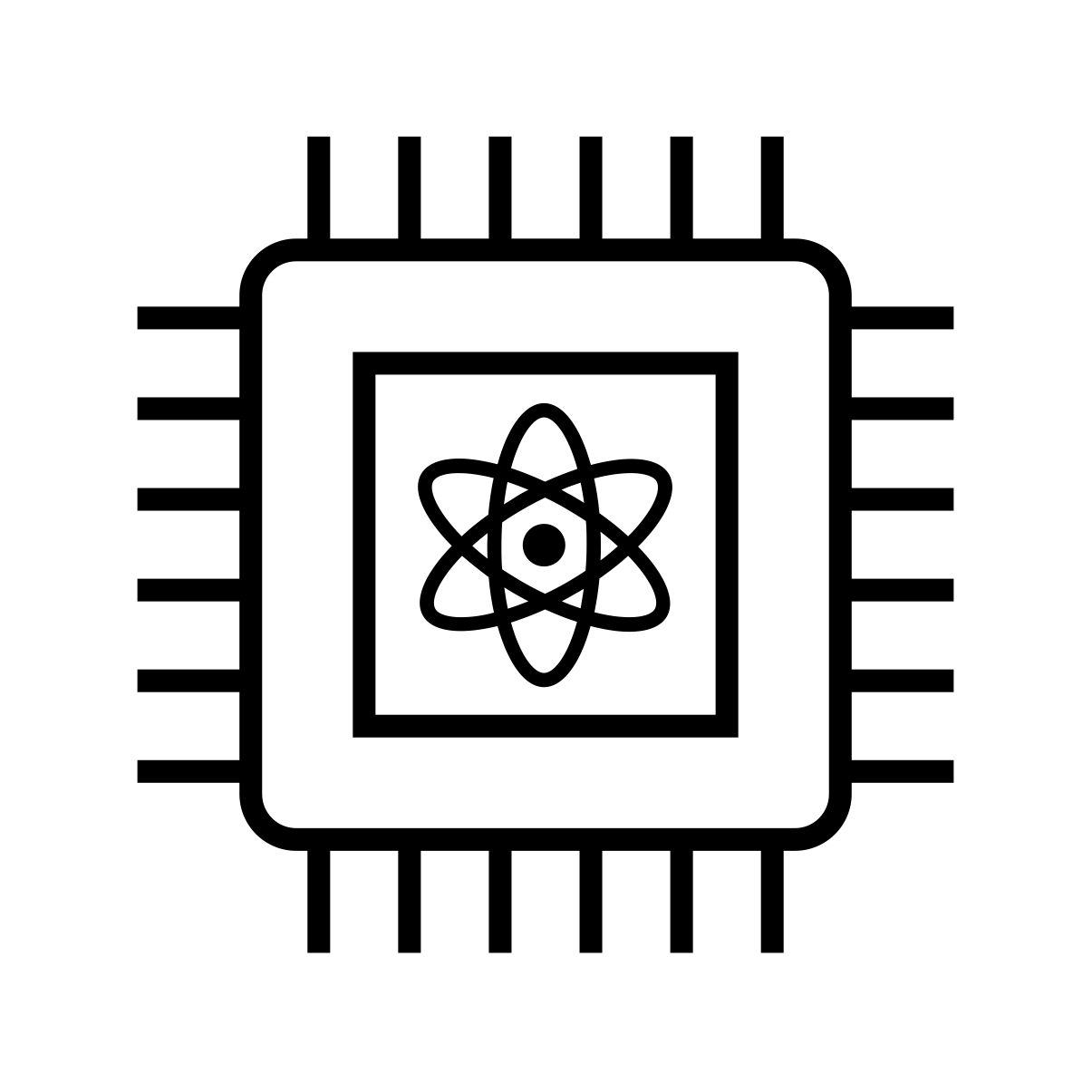}};
        \node (energy) [draw, rounded corners, fill=blue!15, left of=_anchor, node distance=1.5cm, text width=5em] {\centering Electronic energy};
        \node (gradient) [draw, rounded corners, fill=blue!15, right of=_anchor, node distance=1.5cm, text width=5em] {\centering Nuclear gradient, Hessian};
        \node[text width=22em] at (8,0.25) {Propagate QM and MM atoms for the next cycle};
            
        \draw[->, line width=2, to path={|- (\tikztotarget)}] (full.south) to (qmmm.west);
        \draw[->, line width=2, to path={|- (\tikztotarget)}] (qmmm.south) to (embed.east);
        \draw[->, line width=2, to path={|- (\tikztotarget)}] (embed.south) to (orbs);
        \draw[->, line width=2] (orbs.south) to (ham);
        \draw[->, line width=2, to path={-| (\tikztotarget)}] (ham.east) to (qsim);
        \draw[->, line width=2, to path={-| (\tikztotarget)}] (qsim.west) to (energy);
        \draw[->, line width=2, to path={-| (\tikztotarget)}] (qsim.east) to (gradient);
        \draw[->, line width=2, to path={|- (\tikztotarget)}] (energy.north) to (full);
        \draw[-, line width=2, to path={|- (\tikztotarget)}] (gradient.north) to (full);

        \begin{scope}[on background layer]
            \node (bk1) [back group] [fit=(full) (qmmm) (embed) (orbs) (ham), label={above:{\hspace{.7cm} Pre/post-processing on conventional HPC}}] {};
            \node (bk1) [back group] [fit=(qsim) (energy) (gradient), label={above:{\hspace{2.0cm} Quantum hardware}}] {};
        \end{scope}
        
    \end{tikzpicture}
    \caption{Multiscale simulation workflow for embedding quantum computational capabilities within a surrounding classical molecular dynamics environment. The workflow consists of several nested layers of abstraction. At the highest level, we identify some molecular target entity within a larger system; while the former is resolved via quantum mechanics (QM) methods, the latter is treated at the classical molecular mechanics (MM) level for computational tractability. Within the QM region, the molecule is further partitioned into an active subsystem and surrounding environment via projection-based embedding (PBE), allowing a subdomain to be treated at a higher level of QM theory, while the environment is rendered at the density functional theory (DFT) level. Finally, within the embedded QM subsystem we may deploy qubit subspace techniques to further reduce the qubit overhead to utilise near-term quantum hardware in large-scale molecular simulation workflows. This allows us to leverage quantum processing units (QPUs) integrated with high-performance computing (HPC) platforms. Sotorasib molecule in water solvent drawn with VMD \cite{VMD}.}
    \label{fig:simulation_workflow}
\end{figure*}

Quantum computers are a natural platform for simulating chemical systems as they can encode quantum states in linear space, rather than the exponential space required by classical devices. So far, they have been successful in realising small-scale demonstrations of electronic structure calculations. The variational quantum eigensolver (VQE) has facilitated simulations of up to 12 qubits \cite{peruzzo2014variational, Shen2017, OMalleyBabbush2016, Santagati2018a, Kandala2017, Colless2018, hempel2018quantum, kandala2019error, Nam2020, Smart2019, McCaskey2019, Rice2021, arute2020hartree, Gao2021a, Kawashima2021Optimizing, eddins2022doubling, Yamamoto2022, Kirsopp2022, huang2022variational, lolur2023reference, leyton2022quantum, liang2023napa, Motta2022, OBrien2022, khan2023chemically, zhao2023orbital, guo2023experimental, weaving2023benchmarking, liu2023performing, dimitrov2023pushing, jones2023precision, liang2023spacepulse, weaving2025contextual, tilly2022variational}, while more recent developments in quantum-selected configuration interaction (QSCI) have unlocked scales up to the $77$-qubit level \cite{robledo2024chemistry, kaliakin2024accurate, liepuoniute2024quantum, barison2024quantum, shajan2024towards, Mikkelsen2024, Sugisaki2024, Yu2025, piccinelli2025quantum}. For the first time in the chemical sciences, simulations in excess of $100$ qubits are within sight. However, all the above works have been limited to gas-phase calculations, either to study small (often diatomic) systems in minimal atomic orbital basis sets, or modest active spaces of larger molecules and/or basis sets.
    In order to reach the milestone of quantum utility in this field, in which quantum computers can produce viable solutions to problems beyond the reach of exact solutions, there must be development in quantum-enhanced simulation of typical chemical workloads which include modelling the effects of large-scale environment regions such as solvents, biomolecules and surfaces.

The hybrid quantum mechanics/molecular mechanics (QM/MM) method is widespread, and allows one to situate a quantum mechanical calculation within a classical medium of point-charges, resolved using molecular mechanics (MM). The possibility of integrating quantum computational resources in QM/MM has been suggested \cite{Blunt2022Perspective, Ma2023Multiscale, capone2024vision, Santagati2024Drug, Weisburn2025, gunther2025How}, but implementations on real quantum devices are scarce and do not make appreciable use of the hardware, nor provide any scalability guarantees \cite{Izsak2023pharma, hohenstein2023efficient, li2024hybrid, Ettenhuber2025Calculating}. It is becoming increasingly common to see quantum processing units (QPUs) being integrated with high-performance computing (HPC) platforms, bringing with it a need for computational workloads that challenge both quantum \textit{and} classical resources. The QM/MM framework provides a realistic route towards achieving large-scale simulations that fully utilise the augmented capabilities of heterogeneous HPC~+~QPU systems \cite{wintersperger2022qpu, beck2024integrating, elsharkawy2024integration, alexeev2024quantum}.

The motivation behind attempting such a computation lies in the exploration of the abilities of quantum computing in chemistry and material science, particularly in assisting calculations on systems which have complex electronic interactions within a large system. Prominent examples include catalysis and biomolecular systems. While current evidence for quantum advantage in stand-alone (i.e. full system) quantum chemistry simulations requires further work \cite{Lee2023}; we believe that a reasonable application of near-term quantum hardware is the deployment of quantum algorithms in a small, highly-correlated region of a much larger classical QM/MM routine.

Sections \ref{sec:solvation}, \ref{sec:embedding} and \ref{sec:subspace} are dedicated to reviewing techniques that can be used to reduce the quantum resources required to simulate a chemical system, many of which we utilise in a proof-of-concept demonstration presented in Section \ref{sec:hardware_poc}. Combined application of the methods discussed within this article allows quantum computation to be deployed within large-scale chemical simulation workflows, providing a practical route to scientific and industrial utility for the technology.

Section \ref{sec:solvation} discusses classical techniques for treating large-scale environmental effects, including QM/MM for general environments and continuum models designed specifically for solvents. In Section \ref{sec:embedding} we review two embedding techniques; projection-based embedding (PBE) is chemically-motivated and allows a QM calculation to be conducted at two different levels of chemical theory, while density matrix embedding theory (DMET) leverages the Schmidt decomposition in a similar vein as tensor networks to embed a subsystem within a surrounding bath. In Section \ref{sec:subspace} we review qubit subspace techniques that exploit (approximate) symmetries for additional resource reduction, such as qubit tapering and the contextual subspace method.

As a proof-of-concept for this approach, we present a workflow in Section \ref{sec:hardware_poc} (also visually outlined in Figure \ref{fig:simulation_workflow}) where a QSCI simulation of the proton transfer mechanism in water, related to the structural debate over the aqueous form of the hydronium ion \ce{[H3O]+} \cite{ResolvingStructuralDebate2021}, is performed. The results were obtained from the IQM \texttt{QExa20} 20-qubit superconducting device, integrated with the HPC cluster SuperMUC-NG at the Leibniz Supercomputing Centre (LRZ). While the QM/MM framework allows for the study of large chemical systems, realistic candidate structures for the QM subdomain are typically too large to be directly solved using quantum computers. One must therefore layer additional quantum embedding and qubit subspace methodologies to further distil the problem for feasibility on near-term quantum hardware.

\section{Classical Chemical Environments}\label{sec:solvation}

Methods to embed a more accurate quantum model within an extended classical region are motivated by prohibitive computational cost for full-system \textit{ab initio} treatment. These methods allow for important classical effects to be included within quantum chemical calculations, such as the presence of complex molecular structures and surfaces, as well as solvation baths. Here we discuss two modes for treating these external environments; an explicit description via quantum mechanics/molecular mechanics (QM/MM), and an implicit description via continuum models.

\subsection{Quantum Mechanics/Molecular Mechanics}\label{subsec:qmmm}

The QM/MM method has emerged as a popular technique for studying many chemical systems where full treatment at the two individual levels of theory is either intractable or insufficient for modelling the desired properties. For example, this will be the case in biological systems where metal-ion coordination complexes contain regions of high electron correlation \cite{Senn2009Biomolecular, Tzeliou2022Metalloproteins}. 
    The method has also been used extensively throughout other fields in chemical/materials simulation, including photochemistry \cite{Boulanger2018photochemistry}, surface chemistry \cite{Hofer2015Combining}, and condensed matter physics \cite{Hunt2016extended}.
Even the use of favourably scaling quantum chemistry methods like DFT on systems like these can be challenging, and classical force-field approaches cannot capture all the relevant electronic effects.

Introduced in 1976 by Warshel and Levitt \cite{QMMM}, the idea of QM/MM is simple: treat a region of chemical interest with a high-accuracy, computationally-expensive quantum chemistry method, and the rest of the system is modelled with the less accurate but computationally cheaper MM. This partitioning clearly relies on the assumption that the targeted electronic effects within the QM region are mainly local and do not depend on long-range interactions with parts of the system within the MM region. 

It is easy to find the individual energies of the QM and MM regions applied at their respective levels of theory, but including the interactions between the two regions is where the crux of implementing QM/MM lies. A simple approach is known as \textit{subtractive} coupling, where the total energy is given by

\begin{equation}
    E^{\mathrm{QM/MM(sub)}}_{\mathrm{full}} := E^{\mathrm{QM}}_{\mathrm{QM}} + E^{\mathrm{MM}}_{\mathrm{full}} - E^{\mathrm{MM}}_{\mathrm{QM}},
\end{equation}

\noindent where the subscripts indicate which region is included in the computation and the superscripts denote the level of theory applied to that region. As the only QM calculation in the subtractive scheme is applied to the QM region, the interactions between the QM and MM regions are treated at the MM theory level. It is easy to implement as the QM and MM codes do not need to communicate with each other, however there are some key disadvantages to this method including the inability to model the effect of polarisation of the QM region by the MM atoms \cite{Senn2009Biomolecular}. The most common example of subtractive QM/MM is the ONIOM method \cite{ONIOM1, ONIOM2}.

To capture some of these additional effects, the \textit{additive} class of coupling methods can be used. In general, these methods consider the three types of interactions separately and sum them to get the total energy, for example

\begin{equation}
    E^{\mathrm{QM/MM(add)}}_{\mathrm{full}} := E^{\mathrm{QM}}_{\mathrm{QM}} + E^{\mathrm{MM}}_{\mathrm{MM}} + E^{\mathrm{QM/MM}}_{\mathrm{full}},
\end{equation}

\noindent where the third term includes the elusive QM/MM couplings \cite{Blunt2022Perspective}. This coupling term can take several forms, most commonly:

\begin{itemize}
    \item \textbf{Mechanical embedding:} the interactions are treated at the MM level, i.e. the usual MM force field parameters are used to model bonds, angles, torsions etc between bonded QM and MM atoms and Lennard-Jones and Coulomb potential terms for non-bonded atoms \cite{Groenhof2013}.
    \item \textbf{Electrostatic embedding:} The QM Hamiltonian includes point charges from the MM environment as one-electron terms in the Hamiltonian, thus allowing one-way polarisability of the QM atoms by the MM atoms (see Eq. \ref{eq:qmmm_ham}).
    \item \textbf{Polarisable embedding:} Both the QM and MM regions are mutually polarisable and are solved in a self-consistent procedure \cite{cao-ryde-2018}. 
\end{itemize}

As an example, the Hamiltonian for the QM region under electrostatic QM/MM coupling (in atomic units) is 

\begin{align}\label{eq:qmmm_ham}
    \hat{H}_{\mathrm{QM/MM}} = 
    & \underbrace{ -
        \frac{1}{2}\sum_{i}^{N_{\mathrm{el}}}{\nabla_{i}^2}+\sum_{i}^{N_{\mathrm{el}}}{v(\mathbf{r}_i)}+\sum_{i<j}^{N_{\mathrm{el}}}{\frac{1}{|\mathbf{r}_{i}-\mathbf{r}_{j}|}}
    }_{\text{\textbf{\textit{a}}}} \nonumber \\
    & \underbrace{ +
        \sum_{A}^{N_{\mathrm{MM}}} \sum_{B}^{N_{\mathrm{QM}}} \frac{Q_{A} Q_{B}}{|\mathbf{R}_{A}-\mathbf{R}_{B}|}
    }_{\text{\textbf{\textit{b}}}} \nonumber \\
    & \underbrace{ -
        \sum_{A}^{N_{\mathrm{MM}}} \sum_{i}^{N_{\mathrm{el}}} \frac{Q_{A}}{|\mathbf{R}_{A}-\mathbf{r}_{i}|}
    }_{\text{\textbf{\textit{c}}}},
\end{align}

\noindent where the three components \textbf{\textit{a}}, \textbf{\textit{b}}, \textbf{\textit{c}}, correspond to the electronic Hamiltonian, QM-MM nuclear repulsion, and QM-MM electronic-nuclear attraction respectively. The other interaction energies related to the MM region are accounted for by the MM driver outside of this Hamiltonian.

    In recent years, explorations of the potential use of quantum computing in QM/MM workflows have appeared in the literature. In a perspective piece by Blunt et al. \cite{Blunt2022Perspective}, the scaling of the fault-tolerant quantum phase estimation (QPE) algorithm was investigated within the context of large-scale biomolecular simulations. Whilst it is noted that algorithmic advances such as qubitization and Hamiltonian truncation have reduced predicted QPE runtimes by orders of magnitude, workflows relevant to pharmaceutical chemistry remain beyond the limits of reasonable predicted resource expectations. It is therefore noted that embedding approaches, specifically QM/MM and the similar QM-cluster approach will be essential for large scale chemical simulations even in the fault-tolerant regime. Similarly, the review by Capone et al. \cite{capone2024vision} offers some ideas on the future of multiscale chemical modelling with quantum computing, where embedding strategies including QM/MM are likely to have an essential role in enabling more chemically-relevant studies using quantum devices in the coming decades. Notwithstanding the many improvements needed in quantum hardware and algorithms, Santagati et al. \cite{Santagati2024Drug} acknowledge the potential for quantum computing to enhance the precision of QM/MM like methods by targeting the quantum computer at the region of highest electron correlation in an extended system. Some simulated-VQE results were presented by Ma et al. \cite{Ma2023Multiscale} which made use of the many-body expansion method, which was chosen for its potential for integration into larger scale QM/MM workflows. In addition, polarisable embedding utilising a VQE subroutine to solve the wavefunction parameters and dipole moments self-consistently is introduced and performed on emulated hardware by Kjellgren et al. \cite{Kjellgren2024variational}, and subsequently emulated with GPU acceleration within \texttt{CUDA-Q}\cite{CUDA-Q}.

    There are also several examples of actual QM/MM workflows being performed on quantum hardware, where constrained active spaces are employed to make the calculations feasible on current devices. Izs{\'a}k et al. \cite{Izsak2023pharma} performed a 4 electron, 4 orbital (4e, 4o) active space energy evaluation of the enzyme ferredoxin hydrogenase and the photosensitizer temoporfin embedded within larger classical environments on Rigetti superconducting hardware, with VQE and iterative QPE algorithms respectively. Li et al. \cite{li2024hybrid} studied a minimal (2e, 2o) active space of five atoms of a cancer target and drug interaction. This work embeds a VQE energy evaluation of this space on superconducting device within classical surroundings. Additionally, an embedded (6e, 6o) active space computation of the proton transfer step in carbonic anhydrase II was performed by Ettenhuber et al. \cite{Ettenhuber2025Calculating} using both trapped-ion and superconducting devices.

    Finally, recent work by Weisburn et al. \cite{Weisburn2025} and G{\"u}nther et al. \cite{gunther2025How} develops a three-level QM/QM/MM embedding scheme, where the two quantum regions are partitioned with bootstrap embedding. Quantum computations in these works are emulated, but they offer a potential path for multiscale chemical calculations with real quantum device assistance.

\subsection{Continuum Models}

Explicit solvent models, which treat solvent molecules individually, can be computationally prohibitive at a quantum mechanical level, particularly for large systems. An efficient alternative is the use of continuum solvation models\cite{tomasi2005}, which describe the solvent as a continuous dielectric medium, thereby significantly reducing computational cost while capturing key solvation effects. A single solute molecule is immersed in an infinite solvent reservoir and treated at a homogeneous QM level. This solute can be a supermolecule composed of multiple molecules, including solvent molecules when appropriate. The use of continuum models enables efficient quantum mechanical calculations of the solute, allowing for the exploration of molecular properties in solution and providing valuable insights into solvent effects on structural stability, energetics and spectroscopy. Two widely used continuum solvation models are the polarisable continuum model (PCM)\cite{cossi2002} and the conductor-like screening model (COSMO)\cite{klamt1993}, both of which are widely used in quantum chemistry and MD simulations for applications such as reaction mechanisms, spectroscopy, and drug design. This section discusses the principles, advantages and applications of these two continuum-based approaches.

The PCM describes the solvent as a polarisable dielectric continuum surrounding the solute\cite{cossi2002}. The solute, typically treated at the quantum mechanical level, is embedded in a cavity constructed within the dielectric continuum. The solvent's response to the solute's charge distribution is modelled typically by solving the Poisson equation to compute the electrostatic potential at the cavity boundary. This potential induces a reaction field, which is incorporated into the solute's Hamiltonian, effectively accounting for solvation effects.

PCM methods are highly versatile and can be applied to a wide range of solvents and solutes. They are particularly effective for studying equilibrium properties, such as solvation free energies, solvent shifts in spectroscopy, electronic excitation energies, and reaction mechanisms in solution. For example, PCM has been employed within a time-dependent DFT to investigate the shifts in absorption and fluorescence energies in passing from apolar to polar solvent\cite{mennucci2012}. While the accuracy of the model depends on the choice of cavity construction and dielectric constant used to model the solvent environment, the flexibility of PCM allows for the inclusion of non-electrostatic contributions, such as dispersion and repulsion interactions, further improving the accuracy\cite{marenich2009}. The computational cost of PCM scales with the complexity of the cavity and the dielectric response, which can be a limitation for large complex systems.

COSMO is another continuum solvation approach that approximates the solvent as a conductor\cite{klamt1993}. The solvent's response is computed by assuming that the dielectric constant of the solvent is infinite. This approximation leads to a significant simplification of the electrostatic equations, improving numerical stability and convergence while keeping it computationally efficient. After the initial conductor-like screening, a scaling factor is applied to account for realistic dielectric effects. COSMO is particularly well-suited for rapid screening of solvation effects in larger molecular systems, such as drug-like molecules or materials. Its efficiency stems from the reduced complexity of the electrostatic problem, which avoids the need for iterative solutions of the Poisson equation.

An extension of COSMO, known as COSMO for real solvents (COSMO-RS)\cite{klamt2005}, integrates the continuum solvent approach and statistical thermodynamics to describe solvent effects beyond electrostatics, making it more predictive for thermodynamic properties. This has made COSMO-RS a particularly useful tool in applications such as solvation energy estimation, partition coefficients, and drug design.

Both PCM and COSMO provide efficient means to model solvation effects without the computational overhead of explicit solvent simulations. PCM is particularly well-suited for high-accuracy quantum chemical calculations due to its rigorous electrostatic treatment, while COSMO and COSMO-RS are advantageous in applications requiring stability and efficiency, such as large-scale screening studies. Recent developments have sought to combine the strengths of both methods, giving rise to hybrid approaches and extensions, such as the conductor-like modification of PCM (C-PCM)\cite{scalmani2010}.

\section{Quantum Embedding Methods}\label{sec:embedding}

Quantum embedding methods enable the use of multiple independent quantum chemistry methods to directly solve a system \cite{EmbeddingMethodsQuantum2020a}. Similarly to QM/MM, a system is partitioned into a region which requires a high level of theory and one which requires only a lower level. In doing so, it is possible to achieve results which are significantly more accurate than the lower level method alone, while being significantly less costly than a global application of the high-level method\cite{EmbeddingMethodsQuantum2020a, FantasyRealityFragmentbased2019}. Implicitly, the Hamiltonian of the composite system is divided into multiple parts 
    $H_{\mathrm{system}} = H_{\mathrm{A}} + H_{\mathrm{B}} + H_{\mathrm{AB}}$
, the Hamiltonian of the individual systems and also the non-additive part $H_{\mathrm{AB}}$. Accurately representing this final term is challenging, and many methods have been developed to do so \cite{AccuracyFrozenDensity2012, AccuracyOrbitalBased2023}. Importantly, there is a great degree of flexibility regarding both the partitioning of the system and the methods applied to each part. Within the context of quantum computing, embedding methods allow for quantum computing resources to be utilised in simulating parts of systems for which the whole would be impossibly large\cite{ScalableApproachQuantum2024}. Clearly this is critical in the NISQ era, in which qubit counts and executable circuit depths are both very restrictive. However, as quantum hardware continues to develop, the flexibility of embedding methods will provide a straightforward path to fully utilise the resources which become available. Further, when fault-tolerant quantum computers are first realised it is likely that (at least initially) they will have few logical qubits and therefore suffer from the same restriction on admissible system size as current NISQ processors\cite{EarlyFaultTolerantQuantum2024}. Quantum embedding methods may again be employed to fully utilise whatever fault-tolerant resources are available. We discuss two embedding methods: projection-based embedding (PBE) and density matrix embedding theory (DMET).

\subsection{Projection-Based Embedding}\label{subsec:PBE}

Originated by Manby et al. \cite{SimpleExactDensityFunctionalTheory2012}, projection-based embedding enables the use of a wavefunction method within DFT. Initially employed with classical methods \cite{ProjectionBasedWavefunctioninDFTEmbedding2019, ProjectorBasedQuantumEmbedding2025}, quantum algorithms can straightforwardly be used \cite{ScalableApproachQuantum2024,QuantumEmbeddingMethod2023}.

The method begins by selecting an \textit{active region} and \textit{environment}. In the original formulation, subsystems are manually selected at the level of atoms \cite{SimpleExactDensityFunctionalTheory2012}, affording flexibility while requiring that chemical intuition is applied. In practice, it can be difficult to predict which selection is appropriate\cite{AccuracyFrozenDensity2012, OrbitalAlignmentAccurate2020} and methods have been developed to perform this automatically\cite{DirectOrbitalSelection2019, OrbitalAlignmentAccurate2020, CorrespondingActiveOrbital}. Having defined a partition, whole-system DFT is performed to obtain a set of optimised molecular orbitals.

Electrons are then localised to the active and environment subsystems using any of a variety of standard localisation procedures such as IBO\cite{IntrinsicAtomicOrbitals2013}, Pipek-Mezey\cite{FastIntrinsicLocalization1989} or SPADE \cite{AutomaticPartitionOrbital2019, ProjectionBasedMolecularQuantum2023}. Virtual orbitals can likewise be localised, for instance via VVO\cite{ValenceVirtualOrbitals2015} or concentric localisation \cite{SimpleEfficientTruncation}. Where multiple system geometries are to be used, localising each individually may lead to changes in the number of active molecular orbitals and a resulting discontinuity in the potential energy surface\cite{OrbitalAlignmentAccurate2020, CorrespondingActiveOrbital}. Procedures have been developed to avoid this, although these naturally involve some compromise between geometries\cite{AutomatedConsistentEvenHanded2023,DirectOrbitalSelection2019, CorrespondingActiveOrbital, OrbitalAlignmentAccurate2020}. 

Having localised electrons into the two regions, we may express the DFT energy in terms of the electron densities of each subsystem $\gamma^{\mathrm{act}}$ and $\gamma^{\mathrm{env}}$ as \cite{AutomaticPartitionOrbital2019}
    
\begin{align}
    E[\gamma^{\mathrm{act}}, \gamma^{\mathrm{env}}] = &\underbrace{\mathrm{Tr}\left(\gamma^{\mathrm{act}} h_{\mathrm{core}}\right) + {g}(\gamma^{\mathrm{act}})}_{\text{energy of isolated active system}}\nonumber\\
    &+ \underbrace{\mathrm{Tr}\left(\gamma^{\mathrm{env}} h_{\mathrm{core}}\right) + {g}(\gamma^{\mathrm{env}})}_{\text{energy of isolated environment system}}\nonumber \\
    &+ \underbrace{{g}(\gamma^{\mathrm{act}}, \gamma^{\mathrm{env}})}_{\text{non-additive two-electron energy}} ,
\end{align}

\noindent where $g$ corresponds to the electron interaction terms, with ${g}(\gamma^{\mathrm{act}}, \gamma^{\mathrm{env}}) ={g}(\gamma^{\mathrm{act}} + \gamma^{\mathrm{env}})  - {g}(\gamma^{\mathrm{act}}) - {g}(\gamma^{\mathrm{env}})$.

Conspicuously absent from the above is the non-additive kinetic term. Having localised the electrons to subsystems, the environment orbitals are projected out of the Hamiltonian, thus suppressing transitions from the active region. The Fock operator of the active region is augmented with a projection term $P_{\mathrm{proj}}^{\mathrm{env}}$, and a term $V_{\mathrm{emb}}$ which includes the mean-field effect of the environment on the active region,

\begin{align}
    {F}_{\mathrm{emb}}^{\mathrm{act}} &= {h}_{\mathrm{core}} + {V}_{\mathrm{emb}}+ {P}_{\mathrm{proj}}^{\mathrm{env}} + {g}(\gamma_{\mathrm{emb}}^{\mathrm{act}}) \nonumber \\
    &= {h}_{\mathrm{emb}} + {g}(\gamma_{\mathrm{emb}}^{\mathrm{act}}).
\end{align}

Two forms of projector are typically used, the $\mu$-shift $({P}_{\mu}^{\mathrm{env}})_{ij} = \mu [{S} \gamma^{\mathrm{env}} {S}]_{ij}$ \cite{SimpleExactDensityFunctionalTheory2012} and Huzinaga ${P}_{\mathrm{huz}}^{\mathrm{env}} = -\frac{1}{2}\big( {F} \gamma^{\mathrm{env}} {S} + {S} \gamma^{\mathrm{env}} {F} \big)$ \cite{ExactDensityFunctional2016}.
Here, $S_{ij}=\braket{\psi_{i}}{\psi_{j}}$ is the overlap of the atomic orbital basis. The effect of $P_{\mu}^{\mathrm{env}}$ is to take the environment orbitals to a high constant energy (typically $10^{6}$), while $P_{\mathrm{huz}}^{\mathrm{env}}$ sends the negative energy levels to their opposite value. Note that the Fermi-shifted Huzinaga projector can account for orbitals with initially positive values \cite{ProjectionBasedCorrelatedWave2018}. The Huzinaga projector is constructed such that it commutes with the Fock operator, as a result, it gives more precise energies \cite{ProjectorBasedQuantumEmbedding2021}. 
    Note that the PBE procedure is performed with only a single-shot embedding, with no need for computationally expensive feedback between classical and quantum methods.

With the environment frozen, the active region is self-consistently optimised under this new Fock operator, returning a wavefunction for the embedded active region $\ket{\Psi_{\mathrm{emb}}^{\mathrm{act}}}$.  Using a selected wavefunction method, which may be run on a quantum device, the energy of the embedded region can be calculated. Taking $H_{\mathrm{emb}} = h_{\mathrm{emb}} + g(\Psi_{\mathrm{emb}}^{\mathrm{act}})$, where the second term is again a two-electron term but now acting upon the embedded active wavefunction. To correct for double counting of the Coulomb term resulting from the environment's effect on the active region, and to negate the energetic effects of projection, a correction term must be included in the final total \cite{AutomaticPartitionOrbital2019},

\begin{align}
    E[\Psi_{\mathrm{emb}}^{\mathrm{act}} ; \gamma^{\mathrm{act}}, \gamma^{\mathrm{env}}] =  &\underbrace{\bra{\Psi_{\mathrm{emb}}^{\mathrm{act}}} {H}_{\mathrm{emb}} \ket{\Psi_{\mathrm{emb}}^{\mathrm{act}}}}_{\text{active region}}\nonumber \\
    &+ \underbrace{E[\gamma_{\mathrm{env}}]}_{\text{environment}} + \ \underbrace{{g}(\gamma^{\mathrm{act}}, \gamma^{\mathrm{env}})}_{\text{non-additive}}\nonumber \\
    &- \underbrace{\mathrm{Tr}\Big( \gamma^{\mathrm{act}} ({V}_{\mathrm{emb}}+{P}_{\mathrm{proj}}^{\mathrm{env}}) \Big)}_{\text{correction}}.
\end{align}

\begin{figure}[t!]
    \centering
    \includegraphics[width=\linewidth]{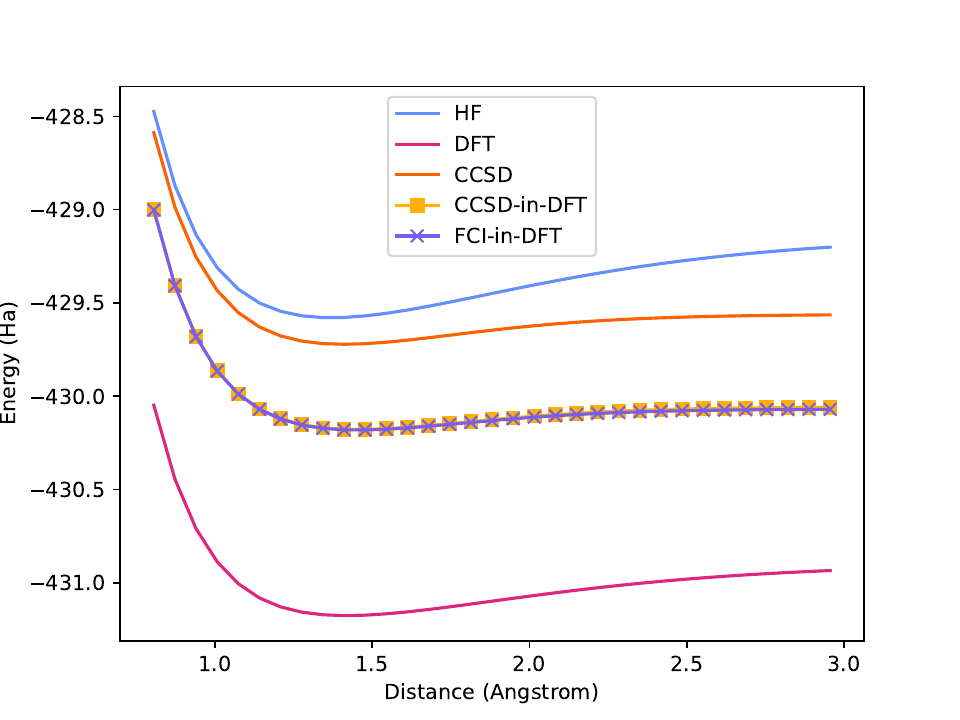}
    \caption{Bond dissociation of perfluoromethane in STO-3G basis. In blue and pink are the whole-system Hartree-Fock and density functional theory (B3LYP). Orange gives the whole system CCSD energy. $\mu$-shift embedded CCSD-in-DFT energy is given by yellow squares, while purple crosses show the embedded FCI-in-DFT energy.}
    \label{fig:pfm}
\end{figure}

The active region wavefunction method may now be run completely independently. Any quantum simulation algorithm can be used, with the other terms derived from classical computation handled as an energy constant. We provide an example of the method in Figure \ref{fig:pfm}, showing the bond dissociation energy of perfluoromethane. In this example, the absolute value of the bond energy is marginally more accurately predicted by DFT alone; however, it serves to illustrate the reduction in problem size achieved by PBE. The \ce{CF3} molecule has 50 spin-orbitals in the STO-3G basis, which is reduced to 28 by embedding.

With currently available quantum hardware, applications have so far been limited to simple demonstrations with only a few atoms\cite{ScalableApproachQuantum2024, QuantumEmbeddingMethod2023}. Quantum advantage is still required to necessitate the use of a quantum processor for the embedded wavefunction method over existing classical methods. We therefore expect quantum-in-classical PBE to become a common technique in the future.

\subsection{Density Matrix Embedding Theory}

Density functional methods such as PBE struggle to elucidate entanglement information between the system and its environment \cite{sun2016quantum}. More sophisticated methods overcome this by replacing the single-particle density with a quantum variable that is better suited to capture entanglement. For example, in condensed matter, Green's function methods such as dynamical mean-field theory are popular \cite{inglesfield1981method, kotliar2006electronic, haule2007strongly, byczuk2008correlated}. However, Green's functions methods often require very large bath spaces to incorporate non-local interactions \cite{held2007electronic}. Together with the difficulty of dealing with time-dependent quantities such as the self-energy, this has limited the application of Green's function methods, particularly in quantum chemistry \cite{wouters2016practical}. Density matrix embedding theory (DMET) as introduced by Knizia and Chan \cite{knizia2012density} was designed to overcome the challenges of Green's functions methods by only dealing with the single-particle density matrix. Furthermore, it was inspired by ideas from tensor networks to efficiently capture entanglement information. 

DMET begins by computing an approximate ground state wavefunction for the full system, $\ket{\Psi_0}$, for example using a truncated configuration interaction theory \cite{booth2015spectral} or anti-symmetrised geminal power wavefunctions \cite{tsuchimochi2015density}. For each subsystem that is to be treated at a higher level of theory a set of bath orbitals are computed from the low-level density matrix. A simple argument based on the Schmidt decomposition of the ground state solution $\ket{\Psi}$ shows that the size of the bath system need be no greater than the size of the subsystem under consideration. Letting $\{S_i\}_{i=1}^{d_S}$ be a basis for the subsystem and $\{B_i\}_{i=1}^{d_B}$ be a basis for the bath we can write

\begin{align}
    \ket{\Psi} &= \sum_{i=1}^{d_S} \sum_{j=1}^{d_B} \psi_{ij} \ket{S_i} \otimes \ket{B_j} \nonumber \\
    &= \sum_{i=1}^{d_S} \sum_{j=1}^{d_B} \sum_{k=1}^{\min{(d_S, d_B)}} U_{ik}\Sigma_{kk} V_{kj}^\dagger  \ket{S_i} \otimes \ket{B_j} \nonumber \\
    &= \sum_{k=1}^{\min{(d_S, d_B)}} \Sigma_{kk} \ket{\tilde{S}_k} \otimes \ket{\tilde{B}_k}.
\end{align}

The set of subsystem and bath orbitals then defines an embedded Hamiltonian to which a high level of theory is applied to obtain the subsystem density matrix $\rho_A$ and energy contribution $E_A$. The global density matrix $\rho = \ket{\Psi_0}\bra{\Psi_0}$ is then optimised self-consistently with respect to some pre-defined cost function designed to match properties of the global state with properties obtained from the collection of subsystems. This process repeats until convergence is achieved. For a full introduction to DMET we refer the reader to Wouters et al. \cite{wouters2016practical}. 
    A related but distinct class of methods exists, known as bootstrap embeddings, which instead enforces consistency by directly matching overlapping fragment density matrices \cite{welborn2016bootstrap, ye2019bootstrap, ye2020bootstrap, meitei2023periodic, tran2024bootstrap}.

In the usual case of classical-in-classical embedding, standard high-level methods such as coupled-cluster theory or the density matrix renormalisation group algorithm can be applied to the embedded Hamiltonian. However, as the size of the active space increases these methods must trade off accuracy and computational tractability. One solution is to treat many subsystems at the higher level of theory although this increases the difficulty of the self-consistent optimisation. An emerging solution is to treat the subsystem on a quantum computer using a quantum algorithm for ground state computation which handles the embedded Hamiltonian. 

As these nascent devices develop they may be able to extend the utility of DMET by allowing for larger active spaces. This idea has been proposed and numerically verified for small systems \cite{rubin2016hybrid, mineh2022solving, li2022toward, cao2023ab}. Additionally, several experiments run on real quantum hardware have yielded results matching classical benchmarks for DMET. Limited to small systems, the combination of DMET with VQE has facilitated simulations of a Hubbard lattice \cite{tilly2021reduced} and hydrogen rings \cite{Kawashima2021Optimizing}. Combining DMET with QSCI allowed a simulation of cyclohexane \cite{shajan2024towards}, which used 32 qubits on a superconducting quantum chip. This proof-of-concept demonstration adds further support for the hope that QSCI may overcome the limitations of VQE; however, it remains clear that further hardware and algorithmic developments will be required before quantum computers are able replace classical methods within the DMET framework. 

    One key limitation for quantum computers is that DMET is iterative and requires the one-particle reduced density matrix of the active space to enforce self-consistency with the environment. While this explicitly captures entanglement information between the active space and environment, often leading to more accurate energy predictions than one-shot embedding methods, it requires computing many more expectation values and performing expensive quantum-classical feedback. This quantum overhead may be partially alleviated by replacing DMET with a bootstrapping embedding as suggested by Liu et al. \cite{liu2023bootstrap}, however the suggested quantum subroutines preclude their implementation on near-term devices.

\section{Qubit Subspace Methods}\label{sec:subspace}

Dimensionality reduction is a common challenge in both conventional and quantum approaches to molecular electronic structure. Methods such as frozen core approximations \cite{AccurateFrozenCore2021a}, active spaces \cite{CorrespondingActiveOrbital} or virtual orbital truncation techniques like Frozen Natural Orbitals (FNO) \cite{sosa1989selection} are commonplace. If used correctly, such methods can retain chemically relevant information at a reduction of computational overhead. These approximations typically rely on physically-motivated assumptions about electronic structure, allowing for systematic removal of orbitals or excitations that are expected to contribute minimally to the correlation energy.

Qubit subspace methods adopt a more abstract approach to yield hybrid orbitals that, while there might be a loss of physical motivation, come with the benefit of additional information being encoded in each qubit via the mixing of molecular orbitals to form an entangled basis in which to describe the molecular system. After application of this orbital-mixing unitary, we apply single-qubit projection operators to fix the state of desired qubits and trace them out of the system. This may be thought of as freezing the new hybrid orbitals, although it also has the flexibility to describe different qubit bases; more generally speaking, we project onto a \textit{stabiliser subspace} of the qubits.

By contrast, in a method such as Complete Active Space Self-Consistent Field (CASSCF), we still apply a unitary to optimise the orbital basis at each step, but the form of this unitary is restricted to single-particle rotations and thus cannot reach as rich a class of molecular orbital bases. In CASSCF, the orbital transformation preserves its class structure, meaning we maintain a strict separation between inactive, active, and virtual orbital spaces, whereas this does not generally apply for qubit subspace methods.

The projection operator for a single qubit indexed $q \in \mathbb{N}$, stabilised by a Pauli operator $P \in \{X,Y,Z\}$, is $\mathbb{P}_{\pm}^{(q)} = \frac{1}{2}[I^{(q)} \pm P^{(q)}]$. For a subset of qubits with indexing set $\mathcal{I} \subset \mathbb{Z}_N$, where $N \in \mathbb{N}$ is the total number of qubits, the projection onto the corresponding stabiliser subspace takes the form 
\begin{equation}\label{eq:qubit_projection}
    \mathbb{P}_{\mathcal{I}, \bm{\sigma}} \coloneqq \bigotimes_{q \in \mathcal{I}} \mathbb{P}_{\sigma_q}^{(q)},
\end{equation}
where $\bm{\sigma}$ denotes the sector that defines the eigenspace.

Moreover, if one wishes to first apply a unitary prior to application of the single-qubit projectors, we may define the rotated projection 
\begin{equation}
    \mathbb{P}_{\mathcal{I}, \bm{\sigma}}^{\prime} \coloneqq U^{\dagger} \mathbb{P}_{\mathcal{I}, \bm{\sigma}} U = \frac{1}{2^{|\mathcal{I}|}} \bigotimes_{q \in \mathcal{I}} [I^{(q)} + \sigma_q U^{\dagger} P^{(q)}U].
\end{equation}
Finally, the reduced qubit subspace is obtained via the map
\begin{equation}
    H \mapsto \mathrm{Tr}_{\mathcal{I}} \big( \mathbb{P}_{\mathcal{I}, \bm{\sigma}}^{\prime} H \mathbb{P}_{\mathcal{I}, \bm{\sigma}}^{\prime} \big),
\end{equation}
where $\mathrm{Tr}_{\mathcal{I}}$ is the partial trace over the qubits indexed by $\mathcal{I}$. We note that, due to the cyclicity of trace operations, one may equivalently view this map as a rotation of the operator $H$, followed by single-qubit projections:
\begin{equation}\label{eq:stabilizer_projection}
    H \mapsto \mathrm{Tr}_{\mathcal{I}} \big( \mathbb{P}_{\mathcal{I}, \bm{\sigma}} U H U^{\dagger} \mathbb{P}_{\mathcal{I}, \bm{\sigma}} \big).
\end{equation}
From an implementation point-of-view, it is typically beneficial to adopt this convention.

This framework is very general, for example if $U$ diagonalises $H$, then this corresponds with solving the problem exactly. Instead, one must design $U$ in such a way that it is classically efficient to realise the rotation $H \mapsto U H U^{\dagger}$, for example by enforcing that $U$ is (near) Clifford. Qubit subspace techniques are differentiated via the way in which we choose the rotation unitary $U$, qubit indices $\mathcal{I}$ and the eigenspace sector $\bm{\sigma}$. In the following subsections we explore several approaches.

\subsection{Frozen Core}\label{subsec:frozen-core}

In quantum chemistry, the canonical molecular orbitals (MO) and corresponding energies are the eigenvectors/values of the Fock matrix, optimised to self-consistency via Hartree-Fock. This is the standard approach to building the second-quantised electronic structure Hamiltonian
\begin{equation}
    H = \sum_{p,q}h_{p,q} a_p^{\dagger} a_q + \sum_{p,q,r,s} h_{p,q,r,s} a_p^{\dagger} a_q^{\dagger} a_r a_s,
\end{equation}
where $h_{p,q}, h_{p,q,r,s} \in \mathbb{R}$ are one and two electron integrals. The Hamiltonian is subsequently mapped onto qubits via a fermion encoding scheme such as Jordan-Wigner \cite{jordan1993paulische} or Bravyi-Kitaev \cite{bravyi2002fermionic}.

\begin{figure}[t!]
    \centering
    \includegraphics[width=\linewidth]{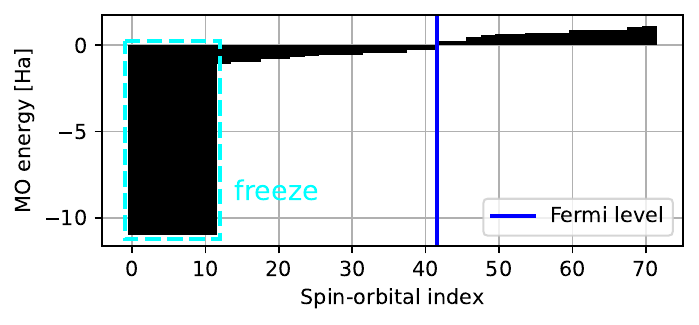}
    \caption{Benzene (\ce{C6H6}) STO-3G molecular orbital energies computed with the restricted Hartree-Fock method. The lowest twelve spin-orbitals may be frozen without dramatically affecting ground-state energy estimates.}
    \label{fig:mo_energies}
\end{figure}

In Figure \ref{fig:mo_energies} we plot the canonical MO energies for benzene, \ce{C6H6}, in a minimal STO-3G atomic orbital basis set. One notes that the core orbitals lie deep with a potential energy well which is typical for most chemical systems. The implication of this is that it is energetically unfavourable for electrons lying deep within this well to be excited into the valence space. Instead, it is more likely that we will observe the greatest electronic activity around the Fermi level, where we find the gap between the highest occupied (HOMO) and lowest unoccupied  (LUMO) molecular orbital. This motivates the notion of a frozen core approximation, in which electrons occupying the lowest-lying MOs are frozen in-place and not allowed to be excited into higher energy states.

Viewed as a qubit subspace technique, this is the simple case where $U = I$ is the identity, and $\mathcal{I} = \{q \in \mathbb{Z}_N: \; \mu_q < -\delta \}$ where $\bm{\mu}$ is the vector of canonical MO energies and $\delta>0$ some energy threshold parameter to truncate the MO orbitals below the core potential. The sector is selected as $\sigma_q = -1\;\; \forall q\in\mathcal{I}$ to enforce that core orbitals are occupied. One may also freeze the valence space in a similar way by instead setting $\sigma_q = +1$ to project onto unoccupied orbitals and truncating the highest-energy MOs, rather than the lowest.

\subsection{Qubit Tapering}\label{subsec:tapering}

\begin{figure}[t!]
    \centering
    \includegraphics[width=\linewidth]{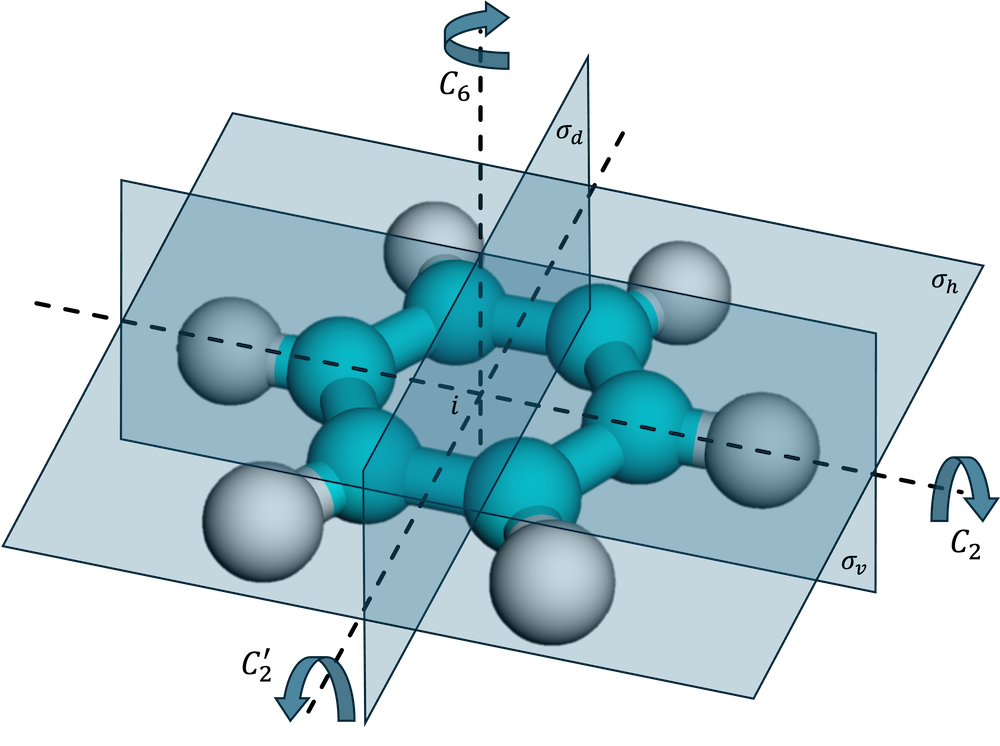}
    \caption{Discrete geometrical symmetries are described by 
        abelian subgroups of
    the molecular point-group, consisting of rotations, reflections and inversions. For example, benzene (\ce{C6H6}) belongs to $D_{6h}$, which consists of group elements $C_6$ ($60^{\circ}$ rotations around the central axis perpendicular to the plane of the molecule), $C_2/C_2^\prime$ ($180^{\circ}$ rotations through axes parallel to the molecular plane), a reflection $\sigma_{h}$ across the horizontal plane, two vertical reflections $\sigma_v/\sigma_d$, and finally the inversion symmetry $i$.}
    \label{fig:benzene_pointgroup}
\end{figure}

In physics, symmetries correspond to conserved quantities in a system of interest\cite{InvarianteVariationsprobleme1918}. Given a Hamiltonian $H$, a symmetry is any operator $S$ such that $[H,S]=0$. For example, in quantum chemistry this may relate to particle number or spin symmetry. The presence of symmetry typically presents opportunities for simplification in some sense, from the exploitation of our knowledge of the problem structure that arises from that symmetry. In chemistry, molecular symmetries guide the construction of better ansatz circuits \cite{gard2020efficient}, or may be used for the purposes of error mitigation through symmetry verification \cite{bonet2018low, sagastizabal2019experimental}. However, one may also exploit physical symmetries as a qubit subspace method.

A subset of symmetries that is of particular interest here are those of $\mathbb{Z}_2$-type, namely operators that describe a form of 2-fold symmetry. The $\mathbb{Z}_2$ symmetries possess a useful property such that, if $[H,S]=0$ and $S$ is $\mathbb{Z}_2$, then $S$ commutes with every term of $H = \sum_k h_k P_k$ individually, i.e. $[P_k, S] = 0 \;\; \forall k$. This fact leads us to a mechanism for reducing the number of qubits in the Hamiltonian without sacrificing any accuracy, since the full and reduced Hamiltonians are isospectral up to a change in eigenvalue multiplicities.

    From the theory of stabilisers
\cite{gottesman1997stabilizer}, given an independent set of $N$-qubit commuting Pauli operators $\mathcal{S}$, there exists a Clifford rotation $C$, a set of qubit indices $\mathcal{I} \subset \mathbb{Z}_N$ and bijective map $f:S \mapsto \mathcal{I}$ such that, for each element $S \in \mathcal{S}$, we have $CSC^{\dagger} = P^{(f(S))}$ for a single-qubit Pauli operator $P \in \{X,Y,Z\}$. In other words, the unitary $C$ maps elements of the set $\mathcal{S}$ onto distinct qubit positions. The positions must be distinct due to the requirement that $\mathcal{S}$ be independent, namely that no single element of the set is a product of other elements, and commuting, so that it is not possible to rotate $X$ and $Z$ onto the same qubit position, say. We may also assume without loss of generality that $P = Z$, since conjugation by Hadamard and phase gates relates the three choices.

Now, suppose we take $\mathcal{S}$ to be the set of $\mathbb{Z}_2$ symmetries of a Hamiltonian $H$. Then, since for each Hamiltonian Pauli term $P_k$ we have $[P_k,S] = 0\;\forall S \in \mathcal{S}$, it must be the case that $[CP_kC^{\dagger},Z^{(q)}] = 0\;\forall q \in \mathcal{I}$ as unitary rotations preserve commutation relations. The implication of this is that the rotated Hamiltonian term $CP_kC^{\dagger}$ must consist of either $I$ or $Z$ in the qubit positions indexed by $\mathcal{I}$. These positions may subsequently be projected and dropped as in Equation \eqref{eq:stabilizer_projection}, or in other words ``tapered'', from the Hamiltonian \cite{bravyi2017tapering}.

When tapering, particular care must be taken to select the correct symmetry sector $\bm{\sigma}$, which can typically be assigned by some knowledge of the underlying problem. In electronic structure, $\mathbb{Z}_2$ symmetries can arise either from spin up/down parity or 
    abelian subgroups of
the geometrical point-group, describing discrete rotations, reflections or inversions of the molecule; an example is given in Figure \ref{fig:benzene_pointgroup}. 
    We note that, while the full point-group might have a high dimensional irreducible representation, the limitation to abelian subgroups greatly restricts the degrees of freedom.
One may refer to point-group tables to correctly choose the desired symmetry sector \cite{setia2020reducing, picozzi2023symmetry}.

\subsection{Contextual Subspace}

In qubit tapering we may only remove as many qubits as there are Hamiltonian symmetries. The goal of the contextual subspace approach \cite{kirby2021contextual, weaving2023stabilizer} is to relax this requirement to permit near-symmetry operators to define the subspace, chosen in such a way that we minimise the loss of information via the projection procedure. As such, this method will introduce some level of systematic error, but if the so-called ``pseudo"-symmetries are selected carefully, this error may be controlled to allow high-accuracy simulations at a considerable saving of qubit resource.

\begin{figure}[t!]
    \centering
    \begin{subfigure}{\linewidth} 
        \centering
        \caption{}
        \includegraphics[width=\linewidth]{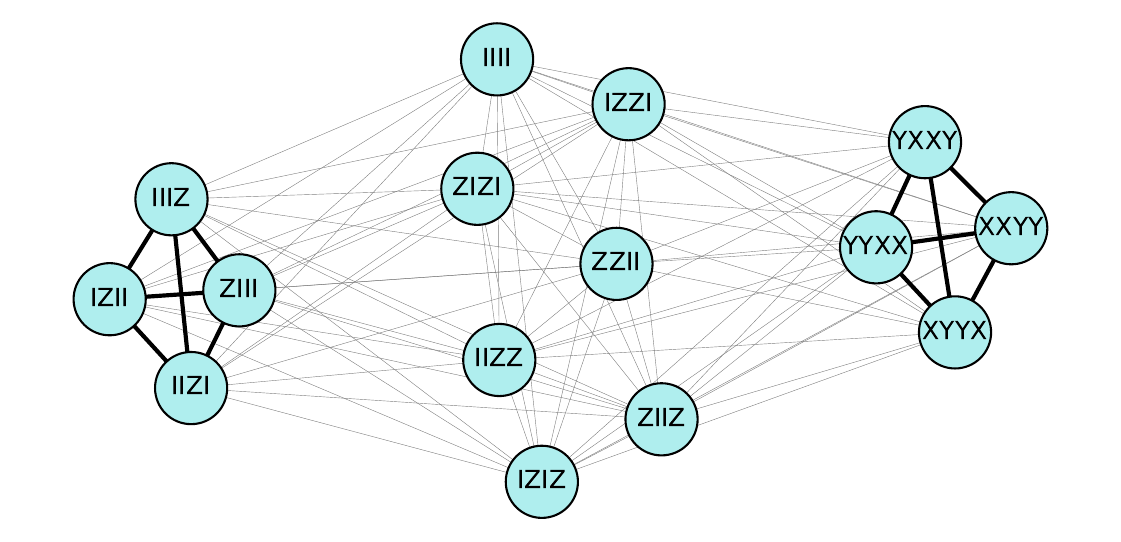}
        \label{fig:H2_noncon_graph}
    \end{subfigure}
    \begin{subfigure}{\linewidth} 
        \centering
        \caption{}
        \includegraphics[width=\linewidth]{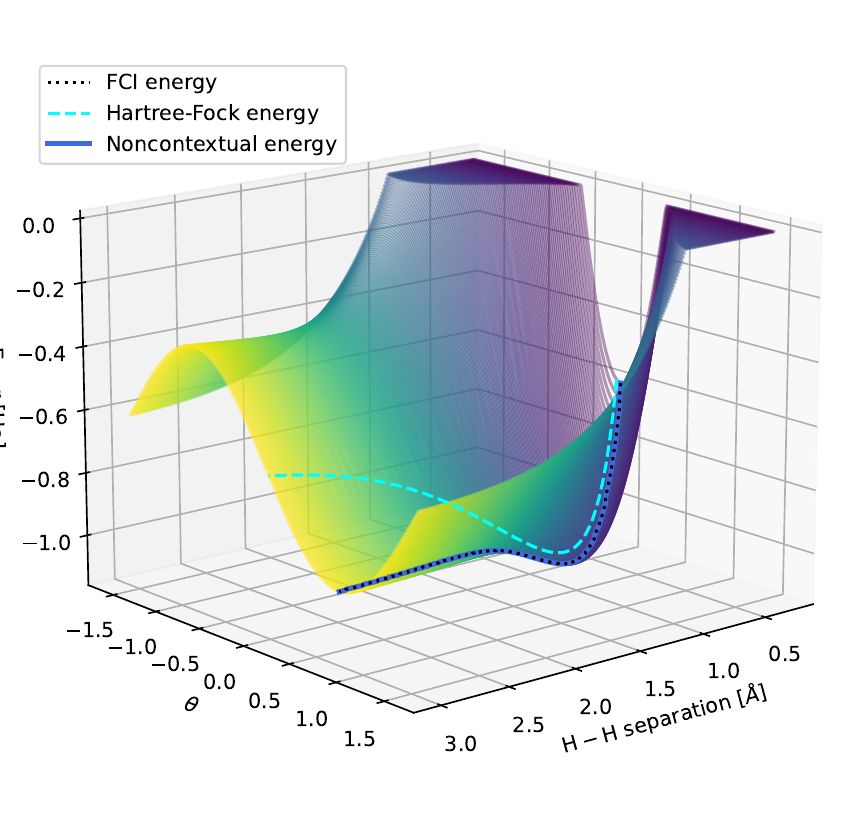}
        \label{fig:H2_noncon_landscape}
    \end{subfigure}
    \caption{Molecular hydrogen, \ce{H2} STO-3G, under the Jordan-Wigner transformation describes a noncontextual system with \textbf{(a)} $2$-clique compatibility graph and \textbf{(b)} noncontextual energy spectrum, whose minimum coincides with the FCI energy.}
    \label{fig:H2_noncon}
\end{figure}

Contextuality provides a broad conceptual picture of non-classical correlation \cite{mermin1990simple, mermin1993hidden, spekkens2007evidence, spekkens2008negativity}; the particular flavour that is exploited in the contextual subspace method is \textit{strong measurement contextuality} \cite{kirby2019contextuality}. In the setting of Pauli operators, this manifests as the presence of measurement contradictions in the same vein as Peres-Mermin magic squares \cite{peres1990incompatible, mermin1990simple}. These contradictions arise from the violation of commutation transitivity among non-symmetry elements of a particular Pauli measurement set. Conversely, a \textit{noncontextual} set is one whose compatibility graph consists of disjoint commuting cliques, all connected to a central symmetry set. An example of a noncontextual Hamiltonian is molecular hydrogen (\ce{H2}) in the minimal STO-3G basis set. Its Pauli terms 
consist of two disjoint cliques $\{ZIII, IZII, IIZI, IIIZ\}$ and $\{XXYY, XYYX, YXXY, YYXX\}$, with the symmetry component $\{IIII, IIZZ, IZIZ, IZZI, ZIIZ, ZIZI, ZZII\}$ as viewed graphically in Figure \ref{fig:H2_noncon_graph}.

A noncontextual Hamiltonian has the structure
\begin{equation}\label{eq:noncon_hamiltonian}
    H_{\mathrm{NC}} = \sum_{P \in \overline{\mathcal{S}}} \bigg(h_{P} + \sum_{k=1}^K h_{P,k} C_k \bigg) P.
\end{equation}
where $\mathcal{S}$ is the symmetry generating set, $K \leq 2N+1$ is the number of disjoint cliques, and $C_k$ is a single representative from each clique, noting that $\{C_k, C_\ell\}=0 \;\forall k \neq \ell$. For example, in Figure \ref{fig:H2_noncon_graph} we have $K=2$, $\mathcal{S}=\{IIZZ, IZIZ, ZIIZ\}$, $C_1 = ZIII$ and $C_2 = XXYY$. One obtains a classical objective function encoding the spectrum of $H_{\mathrm{NC}}$ via a phase-space description of the underlying hidden variable model \cite{Spekkens2016, kirby2020classical, raussendorf2020phase}:
\begin{equation}\label{eq:noncon_objective}
    \eta(\bm{\nu}, \bm{r}) = \sum_{P \in \overline{\mathcal{S}}} \bigg(h_{P} + \sum_{k=1}^K h_{P,k} r_k \bigg) \prod_{S \in \mathcal{S}_p} \nu_S.
\end{equation}
where $||\bm{r}||=1$ and $\mathcal{S}_P \subset \mathcal{S}$ satisfies $P = \prod_{S \in \mathcal{S}_P} S$. Optimising over the parameters $\bm{\nu}, \bm{r}$ yields the noncontextual ground state. In Figure 
\ref{fig:H2_noncon_landscape} we show the $\bm{r}$ optimisation landscape for \ce{H2}, noting that the minimum coincides exactly with the FCI energy.

Molecular hydrogen is a special case; in general, the electronic structure Hamiltonian is dominated by contextual interactions. In the contextual subspace approach, one projects symmetries $\mathcal{S}^\prime \subset \mathcal{S}$ of a noncontextual model system over the full Hamiltonian, with the sector $\bm{\sigma}$ as in the introduction to Section \ref{sec:subspace} identified by optimising the noncontextual objective function of Equation \eqref{eq:noncon_objective}, taking the elements $\nu_S$ that correspond with the chosen symmetries $S \in \mathcal{S}^\prime$. The pairwise anticommuting clique representatives are moreover rotated onto a single Pauli operator prior to the stabiliser subspace projection, either via the sequence of rotations ($R_{\mathrm{SeqRot}}$) or linear-combination of unitaries ($R_{\mathrm{LCU}}$) construction \cite{ralli2023unitary}. Paired with the same Clifford rotations as in qubit tapering that map $\mathcal{S}^\prime$ onto single-qubit Pauli operators, i.e. $CSC^\dagger=Z^{(f(S))}$ for each $S \in \mathcal{S}^\prime$, we obtain the unitary $U = CR_{\mathrm{SeqRot/LCU}}$ that one applies to the Hamiltonian prior to the single-qubit projections of Equation \eqref{eq:stabilizer_projection}. We note that, due to the $R_{\mathrm{SeqRot/LCU}}$ rotation, the unitary $U$ is non-Clifford; however, in the $LCU$ construction it is guaranteed that the increase in the the number of Hamiltonian terms $L\in\mathbb{N}$ is $L^K$, and even this worst-case increase may only be encountered in highly contrived scenarios \cite{ralli2023unitary}. 

\begin{figure}[t!]
    \centering
    \includegraphics[width=\linewidth]{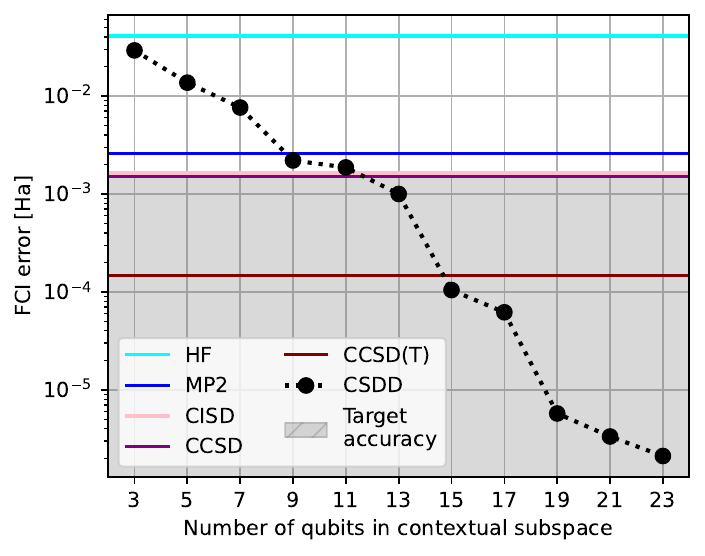}
    \caption{\ce{NaCl} STO-3G FCI errors at $3.5$ \AA ($\approx1.58$ times the equilibrium bond length) against the number of qubits in an expanding contextual subspace. The reduced contextual subspace Hamiltonians were solved via direct diagonalisation (CSDD).}
    \label{fig:csdd_nacl}
\end{figure}

The benefit of the contextual subspace method over those presented above, such as qubit tapering, is that we may project into an arbitrarily small qubit subspace, whereas tapering is limited by the number of $\mathbb{Z}_2$ symmetries present. This feature allows the contextual subspace approach to accommodate any given quantum device, regardless of size, and as such describes a highly flexible approach to quantum computing for many application domains, of which quantum chemistry has been investigated most extensively \cite{kirby2021contextual, weaving2023stabilizer, ralli2023unitary, weaving2023benchmarking, liang2023spacepulse, kiser2024contextual, weaving2025contextual}. Of course, unlike tapering, we accumulate more error as we project into smaller subspaces, so this consideration needs to be navigated carefully. In Figure \ref{fig:csdd_nacl} we demonstrate the decay of error as a contextual subspace is expanded from 1 to 23 qubits for sodium chloride (\ce{NaCl}) STO-3G; while it begins as a 36-qubit problem in its entirety, application of the contextual subspace method allows us to achieve the target accuracy of $1.6$~mHa for just 13 qubits, a considerable saving compared with the full system. These results were obtained using the Python packages \texttt{PySCF} \cite{sun2018pyscf} and \texttt{Symmer} \cite{symmer2022}.

\section{Proof-of-Concept Demonstration}\label{sec:hardware_poc}

In Figure \ref{fig:simulation_workflow}, we outlined a QM/MM simulation workflow for using quantum computational resources for a small region of a molecule, embedded within a larger framework comprising traditional quantum and classical computational chemistry methods. In this section we present results from an initial proof-of-concept example of this workflow where the proton transfer mechanism in water is considered.

\begin{figure}[t!]
    \centering
    \includegraphics[width=\linewidth]{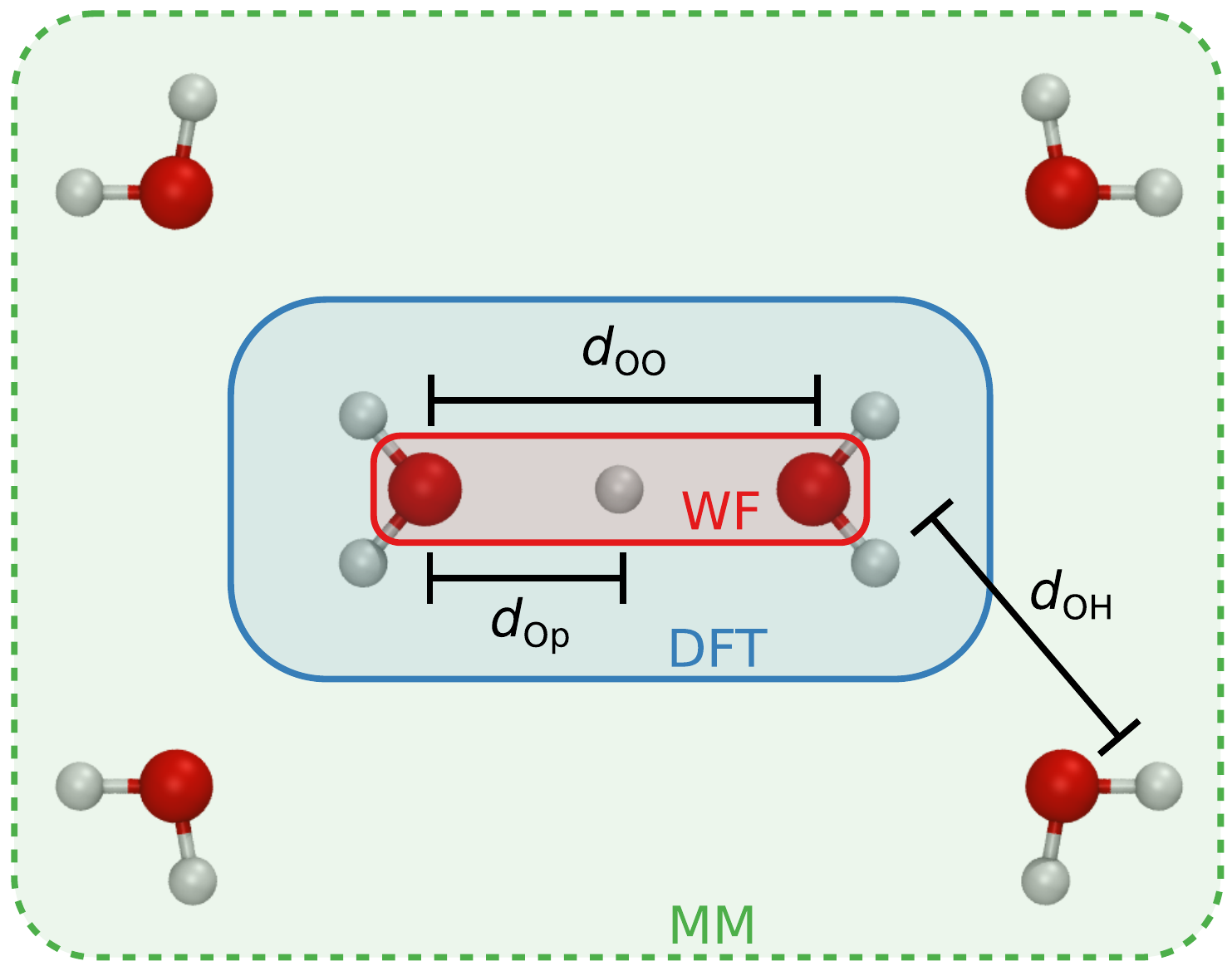}
    \caption{The proton transfer geometry considered in the hardware proof-of-concept experiment. Two water molecules are separated by a distance $d_{\mathrm{OO}}$ Å, with a free proton placed in-between at a distance of $d_{\mathrm{Op}}$ Å from the left water. The proton ratio is defined by $r := d_{\mathrm{Op}}/d_{\mathrm{OO}} \in (0,1)$. These atoms are treated at the QM level, where projection-based embedding is used to split the system into a density functional theory (DFT) and wavefunction (WF) subsystems. Further water atoms are placed around this system and treated at the molecular mechanics (MM) level, with the first solvation shell of four waters placed explicitly at $d_{\mathrm{OH}}$ Å along the OH bonds of the QM waters. The graphic was produced with VMD \cite{VMD}.}
    \label{fig:hydronium}
\end{figure}

\begin{figure}[t!]
    \centering
    \includegraphics[width=\linewidth]{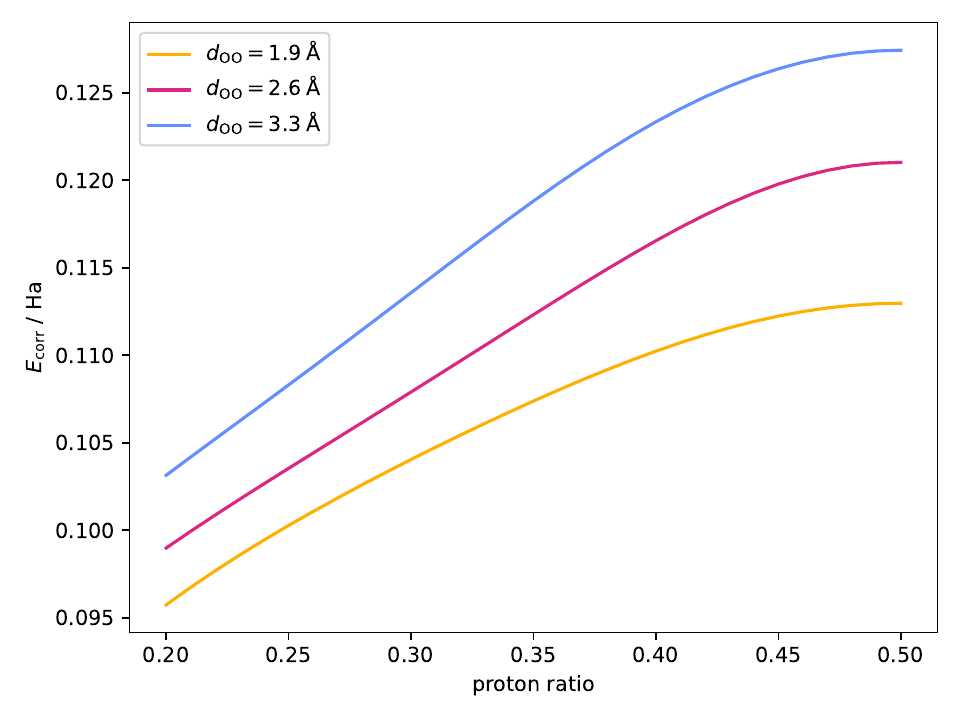}
    \caption{The correlation energy $E_{\mathrm{corr}} = E_{\mathrm{HF}} - E_{\mathrm{FCI}}$, as a function of proton ratio $r$. Restricted-HF and full-CI calculations are performed on a single Zundel cation (the atoms in blue region of Figure \ref{fig:hydronium}) at three water separation distances with the STO-3G basis set. Across the three water separations, the mean correlation energy increase from $r =$ 0.2 to 0.5 is 21\%.}
    \label{fig:correlation_energy}
\end{figure}

Hydronium is the protonated cation of molecular water, i.e. \ce{[H3O]+}. In solution, the simplest hydration structure of hydronium is the Zundel cation \ce{[H5O2]+}, where the hydronium ion is hydrogen-bonded to one other water molecule. For our testbed system we consider a planar Zundel cation, where the distance between the two waters is controlled by a parameter $d_{\mathrm{OO}}$, and the free proton is placed somewhere along this length at a distance $d_{\mathrm{Op}}$ thereby defining a proton ratio $r := d_{\mathrm{Op}}/d_{\mathrm{OO}} \in (0,1)$. Additional solvent water molecules are included around this Zundel system; initially four are placed at a distance $d_{\mathrm{OH}}$ from the four hydrogens of the Zundel cation, and more can be placed randomly in space beyond this. See Figure \ref{fig:hydronium} for a diagram of this geometry. As the free proton moves from being local to one water molecule ($r \approx 0.2$ or $0.8$), to being shared between two ($r \approx 0.5$), the amount of correlation energy in the system increases by about $20\%$ as can be seen in Figure \ref{fig:correlation_energy}.

This system is then partitioned into subsystems which are treated at different levels of theory. The quantum region is defined as the Zundel cation, and the classical region comprises the remaining water molecules. The classical region is treated at the molecular mechanics level, whereas the quantum region is further partitioned into active and environment subregions via projection-based embedding (PBE), described in Section \ref{subsec:PBE}. The latter is treated with density functional theory, while the active region may be treated with conventional post-Hartree-Fock {\em{ab initio}} methods like coupled-cluster, or by constructing a qubit Hamiltonian and evaluating the energy with a quantum algorithm.

For the simulation we set up an electrostatically-coupled QM/MM simulation of the solvated Zundel system via the \texttt{PySCF} qmmm module \cite{sun2018pyscf}, where \texttt{LAMMPS} \cite{LAMMPS} drives the MM force evaluations and the MolSSI Driver Interface ({\texttt{MDI}) package \cite{MDI} facilitates communication between \texttt{PySCF} and \texttt{LAMMPS}. The PBE procedure is executed with \texttt{Nbed} \cite{NBed} using the SPADE localisation method \cite{AutomaticPartitionOrbital2019, ProjectionBasedMolecularQuantum2023}. For the quantum chemistry calculations, the minimal STO-3G atomic basis set is employed, and Kohn-Sham DFT with the B3LYP exchange-correlation functional is used on the environment region. We obtained a Pauli Hamiltonian with \texttt{Symmer} \cite{symmer2022} via the Jordan-Wigner (JW) fermion encoding \cite{jordan1993paulische}.

The resultant qubit Hamiltonian of the core \ce{O-H-O} atoms encodes the exact electronic energy embedded within the DFT environment (for the given basis set and exchange-correlation functional used), with additional polarisation effects caused by the MM atoms which are modelled as point charges (see Section \ref{subsec:qmmm} for further details). We refer to this quantity as the \textit{FCI-in-DFT-in-MM} energy. It is useful to note that the Zundel cation has 15 molecular orbitals in the minimal STO-3G basis set, which would therefore require 30 qubits to describe the QM system on a quantum computer under the JW encoding. However, through the application of PBE (see Section \ref{subsec:PBE}), we may project out the four environment hydrogen atoms, each contributing a single occupied molecular orbital in this basis set and thus resulting in a reduction by 8 qubits in the resulting qubit Hamiltonian. The planar Zundel system has several planes of symmetry, allowing for an additional 4 qubits to be tapered out \cite{bravyi2017tapering}. Finally, we use the frozen-core approximation which brings the final qubit count for our embedded system down to 16 qubits (see Section \ref{sec:subspace} for further detail on these methods).

For the QM/MM simulation, we place an additional 96 water molecules randomly around the Zundel system and perform a molecular dynamics (MD) propagation of these atoms, whilst the coordinates of the QM atoms with a specified $(d_{\mathrm{OO}},\ r)$ parameter set remain fixed. 
    Embedded qubit Hamiltonian data is recorded for 6 steps at a time interval of 2.0fs per step. The system size and number of steps have been chosen to produce a minimally-sized simulation which can be used as a test-bed for the multiscale embedding and subspace methods which are the main focus of this perspective. After all steps have been performed, an averaged Hamiltonian for that parameter set is produced by calculating the mean Pauli coefficient for each Pauli string in the set of Hamiltonians.
The set of mean Hamiltonians are then passed to the IQM superconducting device for the FCI-in-DFT-in-MM energy evaluations.

The energy of the quantum region is calculated on the 20-qubit IQM quantum chip \texttt{QExa20} using quantum-selected configuration interaction (QSCI) \cite{robledo2024chemistry, kaliakin2024accurate, liepuoniute2024quantum, barison2024quantum, shajan2024towards, Mikkelsen2024, Sugisaki2024, Yu2025}. A set of electron configurations (determinants) $\mathcal{D} = \{\ket{\Phi_k}\}_{k=0}^{K-1}$ is sampled from the quantum device. We then form the configuration subspace projection operator $\mathbb{P} \coloneqq \sum_{k=0}^{K-1} \ket{\Phi_k}\bra{\Phi_k}$ and project the electronic structure Hamiltonian into this space
\begin{equation}
    H \mapsto \mathbb{P}H\mathbb{P}=\sum_{k,\ell=0}^{K-1} \bra{\Phi_k}H\ket{\Phi_\ell}\;\ket{\Phi_k}\bra{\Phi_\ell}.
\end{equation}
It is then possible to diagonalise the $K\times K$ matrix $\bm{H}$ with entries $\bm{H}_{k\ell} = \bra{\Phi_k}H\ket{\Phi_\ell}$ using classical compute resources; by solving $\bm{H}v_j=\epsilon_j v_j$ we obtain eigenstates $\ket{\Psi_j} = \sum_{k=0}^{K-1} v_{jk} \ket{\Phi_k}$ of the projected Hamiltonian that satisfy $\mathbb{P}H\ket{\Psi_j} = \epsilon_j \ket{\Psi_j}$. For the hardware results in Figure \ref{fig:hardware-results} we allowed a shot-budget of $10^5$, which produced $K<500$ unique configurations for each simulation, requiring only modest compute resources to diagonalise. Since the electronic energy is obtained via classical diagonalisation, QSCI is more robust to hardware noise than algorithms such as VQE, where the energy estimation itself is susceptible to corruption by noise. By contrast, in QSCI it is only the quality of the configuration subspace that suffers.

The QSCI method requires preparing a quantum state on the quantum chip that approximates the ground state solution or, more generally, shares sufficient support with the ground state to provide an accurate approximation to its energy. The best way to prepare such a state remains an open question. Many previous works fix the parameters of a local unitary cluster Jastrow (LUCJ) ansatz using excitation amplitudes obtained from coupled-cluster calculations \cite{robledo2024chemistry, kaliakin2024accurate, liepuoniute2024quantum, barison2024quantum, shajan2024towards}. For this proof-of-concept demonstration, we instead use a direct matrix product state (MPS) to circuit mapping to warm-start the quantum chip with a low-bond-dimension DMRG solution. We note there is no evidence that either of these approaches will be sufficient for quantum advantage without additional work being done by the quantum device. This is beyond the scope of this article and will be addressed in later work; instead, it serves here as a validation of our QM/MM approach and a proof-of-concept that quantum computational resources may be deployed within large-scale chemical workflows.

The results of the quantum energy evaluations are presented in Figure \ref{fig:hardware-results} alongside the DMRG energy and the HF energy (obtained by direct computation, $\bra{\psi_{\mathrm{HF}}}H\ket{\psi_{\mathrm{HF}}}$). The DMRG energy corresponds to the DMRG solution with bond-dimension truncated to $2$ that is prepared on the quantum chip. We do not claim that our results outperform DMRG in general; indeed, with a moderately high bond-dimension DMRG can be used to find the exact ground state solution for this system. However, the QSCI results obtained from this reference state improve upon its accuracy which motivates the use of this method in the regime where exact DMRG becomes intractable. 
    This accuracy improvement is attributed to the fact that the QSCI energy calculation is optimal within the subspace spanned by the sampled configurations whereas the DMRG energy is only optimal within the manifold of accessible MPS states with a capped maximum bond dimension. In this demonstration, it may also be the case that device noise enables us to sample from a larger set of relevant configurations than are present in the DMRG solution. While this effect of noise has been highlighted elsewhere as a potential benefit to QSCI \cite{piccinelli2025quantum}, we do not expect it to provide a reliable mechanism for larger systems and encourage the development of methods that expand the available configuration space in a more systematic way \cite{Mikkelsen2024, Sugisaki2024, Yu2025}.

\begin{figure}[t!]
    \centering
    \includegraphics[width=\linewidth]{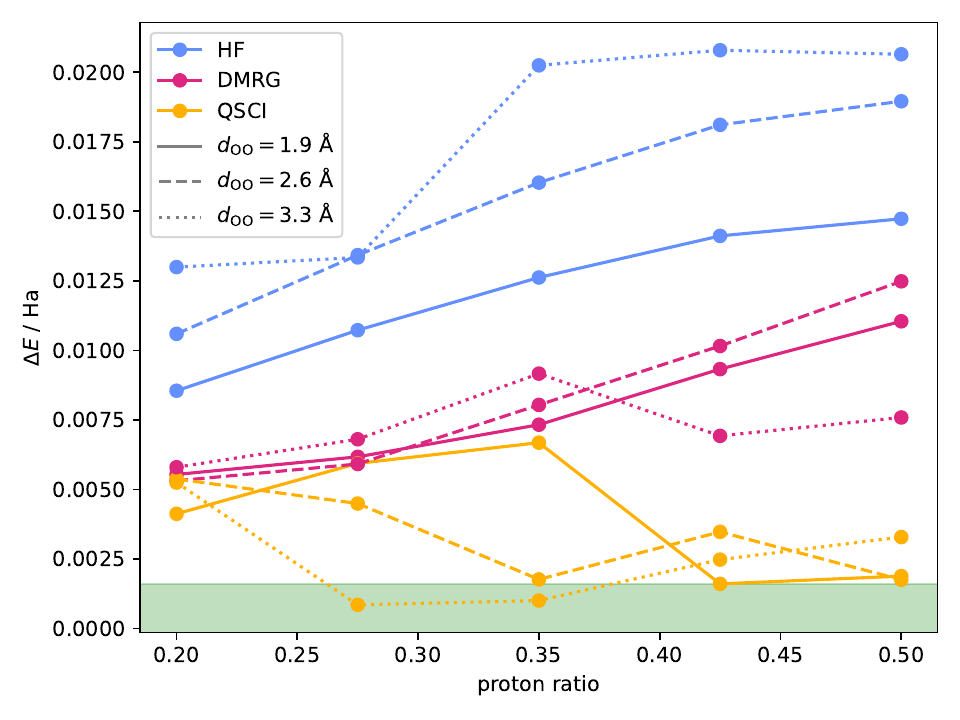}
    \caption{Energy errors of the Zundel system embedded region at different \ce{O-O} separations (indicated by the line style) and proton ratio values, evaluated with three different methods: HF and DMRG use purely classical compute resources, whereas the QSCI energies are found via sampling from the IQM device. All energies are displayed relative to the exact FCI-in-DFT-in-MM energy. The green region indicates the target accuracy of within 1.6mHa to the exact energy. The proton ratio is relative to the \ce{O-O} separation, for example a ratio of 0.5 indicates $d_{\ce{OH}} = d_{\ce{OO}}/2$, i.e. the proton is equidistant between two oxygens.}
    \label{fig:hardware-results}
\end{figure}

The energies were evaluated at three different oxygen separation values, indicated by the line dashing, and at five different proton ratios. For this proof-of-concept demonstration, the Hamiltonian has been chosen such that it can be evaluated with exact diagonalisation, so we report the energies relative to this reference. We also mark the region known as chemical accuracy (within 1.6mHa to this reference), which is motivated by the limits of real-world experimental precision \cite{chemical-accuracy}. 
    However, since we are using the minimal STO-3G basis set, it may be more suitable to describe this region as representing algorithmic accuracy, that is, accuracy within the limits of the basis set as opposed to true quantitative chemical accuracy \cite{review_Motta, MQC}. STO-3G is almost always employed as the default basis in quantum chemistry studies on quantum computers, but as the quality of quantum hardware increases we expect that more effort will be made to move to larger basis sets where results can be considered more chemically meaningful.

As can be seen, the QSCI energies are comfortably within this region at two points in the $d_{\mathrm{OO}} = 3.3$Å evaluation, and near the boundary at several more evaluations. Whilst this precision to the exact solution is not seen across all of our energy evaluations, it is true that our implementation consistently yields more accurate energies than the HF and DMRG direct computation results, the latter of which was used to prepare the quantum circuit which is sampled from.

The proof-of-concept demonstration presented here combines classical molecular dynamics, quantum embedding and subspace methods into a single workflow, which allows quantum energy evaluations of chemical systems beyond the gas-phase regime.

In order to extend this workflow beyond the current fixed quantum geometry scheme, it is necessary to implement evaluations of the energy gradient with respect to nuclear coordinates of the QM nuclei. Within the PBE formalism, analytical nuclear gradients can be computed by introducing a total energy Lagrangian and finding its derivative when minimised with respect to the molecular orbitals \cite{pbe_gradients, Csoka2024Development}. 
    Various quantum algorithms have been proposed for evaluating energy gradients on quantum hardware, both analytical \cite{OBrien2019Calculating, OBrien2022Efficient, Lai2023Accurate} and numerical \cite{Delgado2021Variational, Sugisaki2022Quantum} with requirements ranging from NISQ-friendly to fault-tolerant. Energy gradients may also be approximated directly with a finite-difference approach, although this raises the quantum overhead substantially due to repeated use of the device. We anticipate that challenges surrounding the evaluation of energy gradients can be tackled with a mixture of finite-difference approaches and algorithmic advances. With the ability to compute gradients, the quantum device can then be used to inform the updates of the quantum region nuclei, which inherently include the effect of the point-charge molecular mechanics atoms under the electrostatic QM/MM embedding.
In such a setup, analysis of chemical properties beyond ground state energies is possible. For example, relaxation of the \ce{O-H-O} atoms in the proof-of-concept example would allow for direct study of the proton transfer over a series of MD steps, allowing for accurate estimations of the energy barrier and consequent estimation of the hopping rate.

\section*{Outlook}

The combination of methods described in this perspective have so far been placed in the context of near-term quantum algorithms and processors, applied to small molecular systems embedded within larger classical environments. The advantage of employing QM/MM as the overarching embedding procedure is that, as previously discussed, it is broadly applicable across many fields, including large-scale biomolecular simulations \cite{Senn2009Biomolecular, Tzeliou2022Metalloproteins}, photochemistry \cite{Boulanger2018photochemistry}, surface chemistry \cite{Hofer2015Combining}, and condensed matter physics \cite{Hunt2016extended}. As we look to the future development of quantum-classical hybrid multiscale models which can tackle systems like these, there are some important considerations for the scaling of such methodologies with quantum hardware improvement.

As quantum processors and error mitigation techniques develop, it will be possible to include larger subsystems within quantum algorithms, and to reduce the number of circuit measurements required to obtain accurate results. However, the hybrid methods we describe are still dependent on reasonable scaling of their components. Although these methods are flexible in how computation is partitioned between classical and quantum processors, the overall method may be constrained by either component. For example, in the case of PBE, a full-system DFT calculation is required to initialise the embedding. Similarly, determination of an initial reference state by DMRG constrains the QSCI procedure. As such, admissible systems are those for which approximate classical method can practically be performed.

On the other hand, circuit measurement and readout of quantum processors pose a challenge in scaling. It is important to scrutinise the computational overhead of quantum algorithms to understand their scalability. For example, a popular prediction for Quantum Phase Estimation (QPE) is it could take 25 hours to compute a single electronic point energy for the P450 Cytochrome molecule, based on the availability of a fully error-corrected device comprised of 500,000 physical qubits each with a very low error rate of 0.001\% \cite{goings2022reliably}. Already, this is somewhat prohibitive for practical use-cases, and furthermore this is only for the calculation of electronic energies – to obtain other ground state properties of interest, such as dipole moments or forces, requires considerable additional overhead in the QPE framework. This is because QPE does not permit direct access to the wavefunction, so further processing must be performed on the QPU itself to extract these quantities. By contrast, in VQE, as long as we can write down an observable for the quantity of interest, it can be estimated with no to little overhead beyond the base electronic energy calculation (assuming there is considerable overlap between the Pauli measurements required for these observables and the Hamiltonian). In QSCI, we gain direct access to the wavefunction owing to the classical diagonalisation step, although this imposes a strict classical limit on the size of configuration subspace we can hope to solve, which is the bottleneck of the algorithm. However, this is a limitation of all CI-based methods; the objective of QSCI is instead to identify higher quality subspaces that lead to more compact representations of the wavefunction and thus allow us to extend the limits of CI applicability in terms of system scale.

Beyond intrinsic algorithmic considerations, the scaling of quantum algorithms is consistently challenged by hardware noise. Device fidelities and coherence times are constantly improving, enabling deeper and more complex quantum circuits to be reliably executed. However, in the pre-fault-tolerant regime, error mitigation and error suppression will always be essential in obtaining useful measurement results from the quantum device. For example, for QSCI error mitigation/suppression techniques are essential to maximise the yield of configurations within the correct symmetry sector. Some techniques such as dynamical decoupling \cite{viola1998dynamical, viola1999dynamical} are widely useful and do not significantly increase the computational overhead as the system scales. On the other hand, for readout error mitigation, while being best practice for any quantum circuit execution, care must be taken in its implementation as naive methods exert an additional exponentially-scaling classical overhead to quantum computations \cite{maciejewski2020mitigation, bravyi2021mitigating}. Finally, algorithm-specific error mitigation schemes such as zero-noise extrapolation \cite{li2017efficient, temme2017error} and probabilistic error cancellation \cite{temme2017error, endo2018practical} each come with their own trade-off between accuracy improvements and computational overhead. As we look to scale implementations of hybrid methods, a well-chosen combination of error mitigation/suppression techniques should improve the reliability of quantum computations without accumulating a prohibitive classical overhead.

The integration of these components introduces additional challenges for hybrid workflows. If QPUs are not tightly integrated into classical HPC resources, communication between devices may well undo any practical benefit of hybridisation. \cite{HowBuildQuantum2025}

Overcoming these scaling challenges will be crucial to the development of hybrid quantum-classical algorithms. This is so that as quantum hardware continues to improve, we can begin to study systems of sufficient complexity to challenge both the available quantum and classical compute resources and obtain scientifically important results beyond the reach of classical computers alone.

\section*{Conclusions}

In this perspective, we have outlined a route towards integrating quantum computing into mainstream scientific computing, particularly within the chemical sciences, by demonstrating how quantum devices can serve as supplementary resources that complement conventional approaches to molecular electronic structure calculations. The earliest demonstrations of quantum advantage for problems of real-world interest are likely to emerge through hybrid quantum–classical architectures, where quantum processing units (QPUs) are tightly coupled with high-performance computing (HPC) systems and play roles analogous to GPU accelerators. This integration is already well underway and holds promise for addressing computationally intensive tasks within multiscale and multiphysics simulation frameworks.

To this end, we highlighted a range of classical techniques that can significantly extend the applicability of current noisy intermediate-scale quantum (NISQ) devices (Sections \ref{sec:solvation}, \ref{sec:embedding}, \ref{sec:subspace}). By identifying subdomains within larger simulations that are suitable for quantum treatment, facilitated by the QM/MM method demonstrated here for proton hopping (Section \ref{sec:hardware_poc}), we can offload the quantum-relevant components to QPUs. Furthermore, we presented a flexible and scalable simulation workflow that leverages many of these techniques in tandem, adapting to available quantum and classical resources. Such algorithmic and architectural advances are critical to realising quantum utility well before the advent of fully fault-tolerant quantum computers.

\section*{Author contributions}
TMB: conceptualisation, investigation, software, writing – review \& editing. AM: conceptualisation, investigation, software, writing – review \& editing. TW: conceptualisation, investigation, software, writing – review \& editing. MWdlB: conceptualisation, investigation, software, writing – review \& editing. SW: writing – review \& editing. MN: software. PS: software. AR: writing – review \& editing. PJL: supervision. MC: supervision. MHV: supervision. LS: supervision. PVC: conceptualisation, supervision, writing – review \& editing.

\section*{Conflicts of interest}
The authors declare no conflicts of interest. 

\section*{Data availability}
The data and scripts supporting the proof-of-concept demonstration are openly available on GitHub (https://github.com/UCL-CCS/extending\_digitaldiscovery\_data) and have been archived on Zenodo (https://doi.org/10.5281/zenodo.17304716). The remaining results are generated with the open-source packages \texttt{Nbed}, \texttt{Symmer}, and \texttt{PySCF}, as cited throughout.

\section*{Acknowledgements}
TMB, AM, MWdlB and TW acknowledge support from the Engineering and Physical Sciences Research Council (EPSRC, grant numbers EP/S021582/1, EP/T517793/1 and EP/W524335/1). TW and MWdlB additionally acknowledge support from CBKSciCon Ltd. MN, PS, MC and MHV acknowledge funding by the Munich Quantum Valley, Sections K5 Q-DESSI and K7 QACI. The research is part of the Munich Quantum Valley, which is supported by the Bavarian state government with funds from the Hightech Agenda Bayern Plus. AR and PJL are supported by the NSF STAQ project (PHY-1818914). PVC is grateful for funding from the European Commission for VECMA (800925) and EPSRC for SEAVEA (EP/W007711/1). We thank LRZ for their support and facilitating access to high performance compute, in addition to the IQM \texttt{QExa20} superconducting device.



\balance
\renewcommand\refname{References}
\bibliography{rsc} 
\bibliographystyle{rsc}

\end{document}